\documentclass[acmtog]{acmart}

\usepackage{booktabs} % For formal tables
\usepackage{multirow}
\usepackage{xr}
\usepackage{wrapfig}
\usepackage{hyphenat}
\usepackage{soul,color}
\usepackage{amsopn}
\usepackage{amsmath}

\usepackage{amssymb}
\usepackage{tabularx}
\usepackage{subfigure}

\definecolor{Red}{cmyk}{0,1,1,0}
\definecolor{Green}{cmyk}{1,0,1,0}
\definecolor{Cyan}{cmyk}{1,0,0,0}
\definecolor{Purple}{cmyk}{0.45,0.86,0,0}
\definecolor{Rosolic}{cmyk}{0.00,1.00,0.50,0}
\definecolor{Blue}{cmyk}{1.00,1.00,0.00,0}
\definecolor{BlueViolet}{cmyk}{0.86,0.91,0,0.04}
\definecolor{NavyBlue}{cmyk}{0.94,0.54,0,0}

\newcommand{\maNote}[1]{}

\usepackage[ruled]{algorithm2e} % For algorithms

\SetAlFnt{\small}
\SetAlCapFnt{\small}
\SetAlCapNameFnt{\small}
\SetAlCapHSkip{0pt}

\acmJournal{TOG}
\acmYear{2022}
\acmVolume{41}
\acmNumber{6}
\acmArticle{1}
\acmMonth{12}

\setcopyright{rightsretained}
\acmDOI{10.1145/3550454.3555451}

\begin{document}
\title{Human Performance Modeling and Rendering via Neural Animated Mesh}	

\author{Fuqiang Zhao}
\orcid{0000-0003-2786-5699}
\affiliation{%
	\institution{ShanghaiTech University}
	\city{Shanghai}
	\country{China}}
\affiliation{%
	\institution{NeuDim Digital Technology (Shanghai) Co.,Ltd.}
	%\city{Shanghai}
	\country{China}}
\email{zhaofq@shanghaitech.edu.cn}

\author{Yuheng Jiang}
\affiliation{%
	\institution{ShanghaiTech University}
	\city{Shanghai}
	\country{China}
}
\email{jiangyh2@shanghaitech.edu.cn}

\author{Kaixin Yao}
\affiliation{%
	\institution{ShanghaiTech University}
	\city{Shanghai}
	\country{China}
}
\email{yaokx@shanghaitech.edu.cn}

\author{Jiakai Zhang}
\affiliation{%
	\institution{ShanghaiTech University}
	\city{Shanghai}
	\country{China}
}
\email{zhangjk@shanghaitech.edu.cn}

\author{Liao Wang}
\affiliation{%
	\institution{ShanghaiTech University}
	\city{Shanghai}
	\country{China}
}
\email{wangla@shanghaitech.edu.cn}

\author{Haizhao Dai}
\affiliation{%
	\institution{ShanghaiTech University}
	\city{Shanghai}
	\country{China}
}
\email{daihzh@shanghaitech.edu.cn}

\author{Yuhui Zhong}
\affiliation{%
	\institution{ShanghaiTech University}
	\city{Shanghai}
	\country{China}
}
\email{zhongyh@shanghaitech.edu.cn}

\author{Yingliang Zhang}
\affiliation{%
	\institution{DGene Digital Technology Co., Ltd.}
	%\city{Shanghai}
	\country{China}
}
\email{yingliang.zhang@dgene.com}

\author{Minye Wu}
\affiliation{%
	\institution{KU Leuven}
	%\city{Shanghai}
	\country{Belgium}
}
\email{minye.wu@kuleuven.be}

\author{Lan Xu}
\affiliation{%
	\institution{ShanghaiTech University}
	\city{Shanghai}
	\country{China}
}
\email{xulan1@shanghaitech.edu.cn}
\authornote{The corresponding authors are Lan Xu (xulan1@shanghaitech.edu.cn) and Jingyi Yu (yujingyi@shanghaitech.edu.cn). }

\author{Jingyi Yu}
\affiliation{%
	\institution{ShanghaiTech University}
	\city{Shanghai}
	\country{China}
}
\email{yujingyi@shanghaitech.edu.cn}
\authornotemark[1]

	\begin{abstract}
        We have recently seen tremendous progress in the neural advances for photo-real human modeling and rendering. However, it's still challenging to integrate them into an existing mesh-based pipeline for downstream applications. In this paper, we present a comprehensive neural approach for high-quality reconstruction, compression, and rendering of human performances from dense multi-view videos. Our core intuition is to bridge the traditional animated mesh workflow with a new class of highly efficient neural techniques. We first introduce a neural surface reconstructor for high-quality surface generation in minutes. It marries the implicit volumetric rendering of the truncated signed distance field (TSDF) with multi-resolution hash encoding. We further propose a hybrid neural tracker to generate animated meshes, which combines explicit non-rigid tracking with implicit dynamic deformation in a self-supervised framework. The former provides the coarse warping back into the canonical space, while the latter implicit one further predicts the displacements using the 4D hash encoding as in our reconstructor. Then, we discuss the rendering schemes using the obtained animated meshes, ranging from dynamic texturing to lumigraph rendering under various bandwidth settings. To strike an intricate balance between quality and bandwidth, we propose a hierarchical solution by first rendering 6 virtual views covering the performer and then conducting occlusion-aware neural texture blending. We demonstrate the efficacy of our approach in a variety of mesh-based applications and photo-realistic free-view experiences on various platforms, i.e., inserting virtual human performances into real environments through mobile AR or immersively watching talent shows with VR headsets.  
    \end{abstract}

	
	
	%
	% End generated code
	%
	
	
\begin{teaserfigure}
	\centering
	\includegraphics[width=1.0\linewidth]{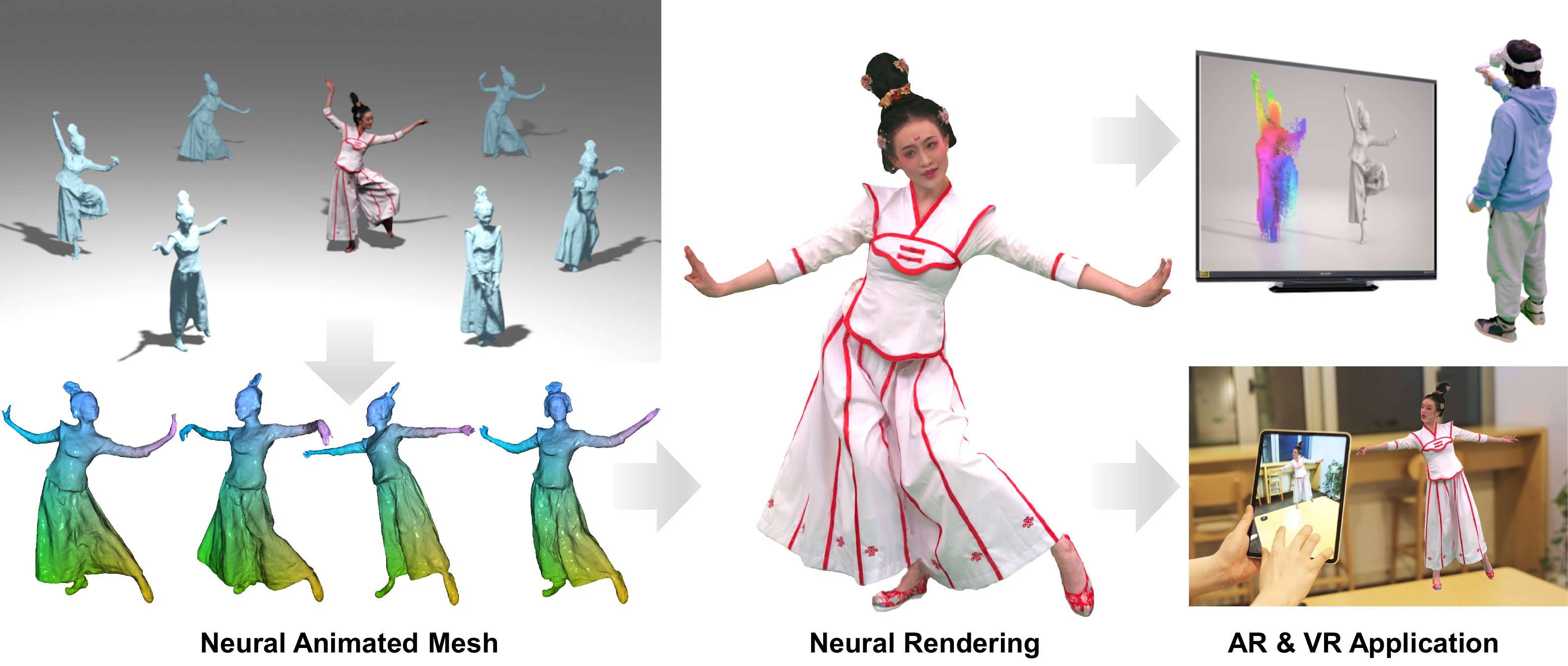}
	\vspace{-21pt}
	\caption{We present a neural human modeling and rendering scheme compatible with conventional mesh-based production workflow. Based on dense multi-view input, our approach enables efficient and high-quality reconstruction, compression, and rendering of human performances. It supports 4D photo-real content playback for various immersive experiences of human performances in virtual and augmented reality.}
	\label{fig:teaser}
\end{teaserfigure}
	
	\maketitle
	
	\section{INTRODUCTION}
We are entering into an era where the boundaries between virtual and real worlds are diminishing. An epitome of this revolution is the volumetric recording and playback of 4D (space-time) human performances that subsequently allows a user to watch the performance immersively in virtual environments. Specifically, a user can interact with the performer as if they were physically present, as simple as moving about to change perspectives and as sophisticated as editing the contents at her fingertips. By far, the most widely adopted workflow to produce volumetric human performance is still to reconstruct and track a dynamic mesh with a per-frame texture map. Significant advances, classic and learning-based, have been made on reconstruction algorithms in both academia~\cite{schoenberger2016sfm,motion2fusion,collet2015high} and industry~\cite{AGISOFT,metastage}, perhaps symbolized by the acquisition of Capturing Reality by Epic Games. In reality, either photogrammetric or 3D scanning-based reconstructions are still time-consuming and remain vulnerable to occlusions and lack of textures which causes holes and noise. Fixing the final reconstructed sequence to meet the minimal immersive viewing requirement demands excessive cleanup works by experienced artists.

Recent neural advances~\cite{NR_survey,NeuralVolumes,Wu_2020_CVPR} attempt to bypass explicit reconstructions and instead focus on synthesizing photo-realistic novel views. Most notably, the Neural Radiance Field (NeRF)~\cite{nerf} replaces the traditional notion of geometry and appearance with a single MLP network where rendering individual pixels in any new camera view maps to per-ray network inference. The original NeRF and its acceleration schemes~\cite{yu2021plenoctrees,mueller2022instant} mainly focus on static scenes whereas more recent approaches ~\cite{peng2021neural,zhao2022humannerf,tretschk2020non,zhang2021editable} aim to extend such neural representations to dynamic scenes with time as a latent variable. Brute-force training and rendering strategies are inefficient and real-time rendering, in particular, requires striking an intricate balance between space~\cite{yu2021plenoctrees,yu_and_fridovichkeil2021plenoxels,mueller2022instant} and quality~\cite{suo2021neuralhumanfvv,wang2022fourier}. In the context of animatable human avatars, it is also possible to produce neural characters by imposing parametric human models as priors ~\cite{realTimeDDC,liu2021neural,bagautdinov2021driving,xiang2021modeling}. Yet, their final quality relies heavily on pre-scanned or parametric templates and tedious time-consuming per-actor training.

% don't support existing mesh-based pipeline
Another major challenge with neural approaches is that their results, i.e., the trained neural networks, do not readily support the existing mesh-based animation pipeline, for either streaming or rendering. Even though it may be possible to extract meshes from the network, e.g., by thresholding the density field from NeRF followed by triangulation, their produced geometry tends to be too noisy to fit into previous mesh-based pipeline~\cite{collet2015high}. The seminal neural implicit surfaces~\cite{wang2021neus,munkberg2021nvdiffrec} substitute the density field with a signed distance field (SDF) in the volume rendering formulation and can recover very high quality geometry. However, these approaches still require long a training time not scalable to generate dynamic mesh sequences. In fact even the mesh sequence can be reconstructed, very few efforts have emphasized applying either neural-based mesh compression to support streaming or neural rendering for high-quality playback. 

% 3. our key idea
We present a comprehensive neural modeling and rendering pipeline to support high-quality reconstruction, compression, and rendering of human performances captured by a multi-view camera dome. We follow the traditional animated mesh workflow but with a new class of highly efficient neural techniques, as shown in Fig.~1. 

% 4. our technical pipeline
Specifically, on mesh reconstruction, we present a highly accelerated neural surface reconstructor, Instant-NSR, analogous to Instant-NGP~\cite{mueller2022instant}, but based on SDF volumetric formulation from NeuS~\cite{wang2021neus} for surface reconstruction. We show that the direct adoption of multi-resolution hash encoding can lead to instability due to gradient explosion at early iterations of feature hashing based network optimization. We instead present a Truncated SDF (TSDF) formulation used in previous 3D fusion frameworks~\cite{KinectFusion,Newcombe2015} implemented via Sigmoid-based mapping. Our open-source PyTorch implementation reduces the training time from 8 hours in NeuS to 10 minutes with comparable quality. We further present a CUDA implementation with finite difference for approximately estimating the normal, which can further reduce the training time but with slightly degraded quality.

% 4.3 tracking: hybrid neural tracking, cuda-based non-rigid tracking + NGP-based delta-motion estimation
We apply Instant-NSR to generate high-quality mesh sequences and then set out to compress the geometry and encode temporal corresponding information using animated meshes. Specifically, we introduce a hybrid neural tracker that combines explicit non-rigid tracking with implicit dynamic deformation in a self-supervised framework. In the explicit phase, we adopt the fast non-rigid tracking scheme~\cite{Newcombe2015,UnstructureLan} based on Embedded deformation~\cite{sumner2007embedded}. It generates coarse initial warping from individual frames to the canonical frame to maintain high efficiency. In the implicit phase, we learn the per-vertex displacements in the canonical space for the 4D coordinates (3D warped position + 1D timestamp) to preserve fine geometric details. We again adopt the 4D hash encoding scheme with CUDA kernels and the volumetric rendering formulation as in Instant-NSR for fast and self-supervised training to obtain the animated meshes.

% 4.4 rendering: to enable streamable downstream application, hierarchical neural blending scheme
For rendering, the brute-force approach would be to adopt dynamic texturing ~\cite{UVAtlas,collet2015high} for supporting streamable immersive experiences. We instead discuss strategies under different bandwidth settings, ranging from using a single texture map across the sequence to streaming videos from all dome cameras for high-quality lumigraph rendering~\cite{buehler2001unstructured}. We show that a compromise of first rendering 6 virtual views covering different perspectives toward the performer and then conducting neural texture blending that accounts for occlusion and view dependency strike an intricate balance between quality and bandwidth. Finally, we demonstrate content playback for various virtual and augmented reality applications, ranging from inserting virtual human performances into real environments on mobile AR platforms to immersively watching ultra high-quality talent shows with VR headsets.  

To summarize, our main contributions include:
\begin{itemize} 
    \setlength\itemsep{0em}

	\item We introduce a novel pipeline to consistently renew the modeling, tracking, and rendering of dynamic human performances into the neural era, favorably compared to existing systems in terms of both efficacy and efficiency.
	
	\item We propose a fast neural surface reconstruction scheme to generate high-quality implicit surfaces in minutes, by combing TSDF volumetric rendering with hash encoding.
	
	\item We propose a hybrid neural tracking approach to generate animated meshes that are compatible with conventional production tools, by combining explicit and implicit motion representations in a self-supervised manner. 
	
	\item We propose a hierarchical neural rendering for photo-realistic human performance synthesis and demonstrate the capability of our pipeline in a variety of mesh-based applications and immersive free-view VR/AR experiences.
	
\end{itemize}

	\section{RELATED WORK}
\paragraph{Neural Human Modeling.} 
Neural implicit representations have recently emerged as a promising alternative to traditional scene representations, such as meshes, pointclouds, and voxel grids.
And due to their continuous nature, neural implicit representations can render images at infinite resolution in theory.
Recently, NeRF \cite{nerf} and followups \cite{pumarola2021d,zhang2021editable,peng2021neural,wang2021ibrnet,zhao2022humannerf, chen2021mvsnerf,wang2021ibutter, wang2021mirror,chen2021snarf,tiwari2021neural,park2021nerfies} utilize volumetric representations and compute radiance by ray marching through a 5D coordinate representations.
Based on the classical volume rendering technique, they achieve impressive results in novel view synthesis.
However, due to the ambiguity of volume rendering \cite{zhang2020nerf++}, these methods still suffer from the low geometric quality.
Some methods~\cite{niemeyer2020differentiable,yariv2020multiview,niemeyer2019occupancy,saito2021scanimate} based surface rendering try to obtain gradients by implicit differentiation, optimizing the underlying surface directly.
UNISURF~\cite{oechsle2021unisurf} is a hybrid method that also learns the implicit surface by volume rendering and encourages the volume representation to converge to a surface.
Different from UNISURF, NeuS \cite{wang2021neus} provides an unbiased conversion from a signed distance function (SDF) into density for volume rendering.
While UNISURF represents the surface by occupancy values and gradually decreases the sample region in some predefined steps to make the occupancy values converge to the surface, NeuS represents the scene by SDF, so the surface can be naturally extracted as the zero-level set of SDF, yielding a better reconstruction accuracy than UNISURF.
What all methods have in common is that they rely on the NeRF training process, i.e., ray march inference on the network during the rendering process, which is computationally expensive during both training and inference.
Thus, they can only apply to static scene reconstruction, but can not be applied to handle dynamic scenes which need unacceptable training time.
Recently, some NeRF extensions \cite{yu_and_fridovichkeil2021plenoxels,wang2022fourier,yu2021plenoctrees,mueller2022instant,zhang2022neuvv} are proposed to accelerate NeRF both on training and rendering, while accelerations for implicit surface reconstruction are missed.
More recently, ~\cite{munkberg2021nvdiffrec} optimizes an explicit mesh representation by utilizing efficient differentiable rasterization with an end-to-end image loss funtion, but still requires hours training.
We aim to propose a fast implicit surface reconstruction method only requiring minutes training with multi-view images.

\begin{figure*}[t]
  \includegraphics[width=\linewidth]{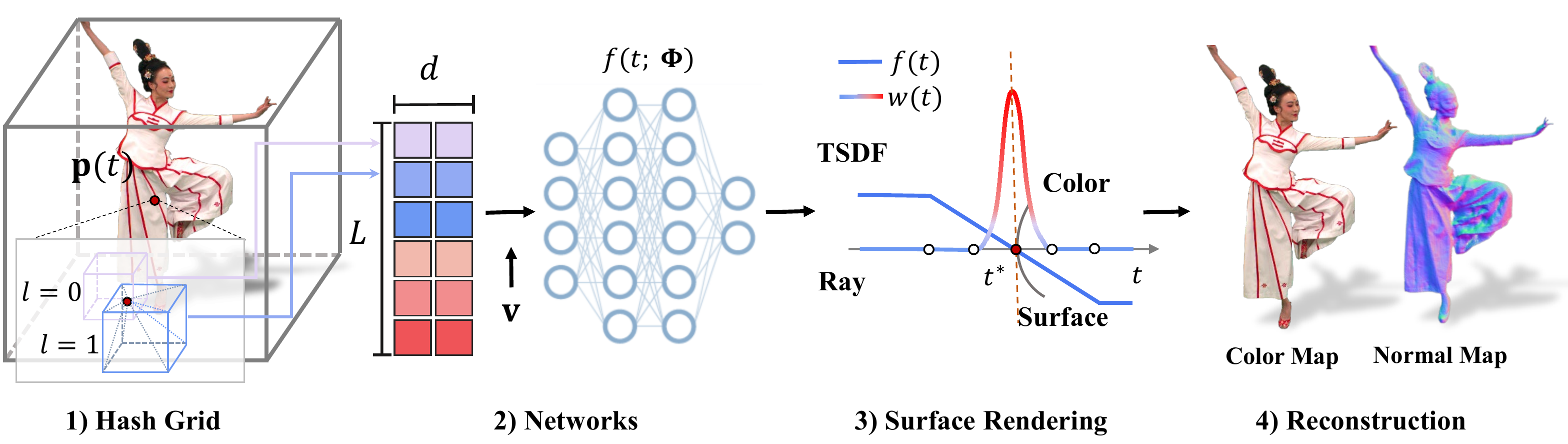}
  \caption{The illustration of our fast surface reconstruction pipeline. (1) for a 3D input position $x$, we first find $L$ surrounding voxels in a 3D multiresolution hash grid and compute $d$ dimension feature with trilinear interpolation of eight corners for each voxel. (2) network takes the generated feature and view direction as inputs, and outputs SDF value and color. (3) utilizes SDF value to compute the weight for blending color of each 3D input position $x$ in a ray, and then generates the final pixel color for photo-metric loss in (4).}
  \label{fig:neus pipeline} 
\end{figure*}

\begin{figure}[t]
  \includegraphics[width=\linewidth]{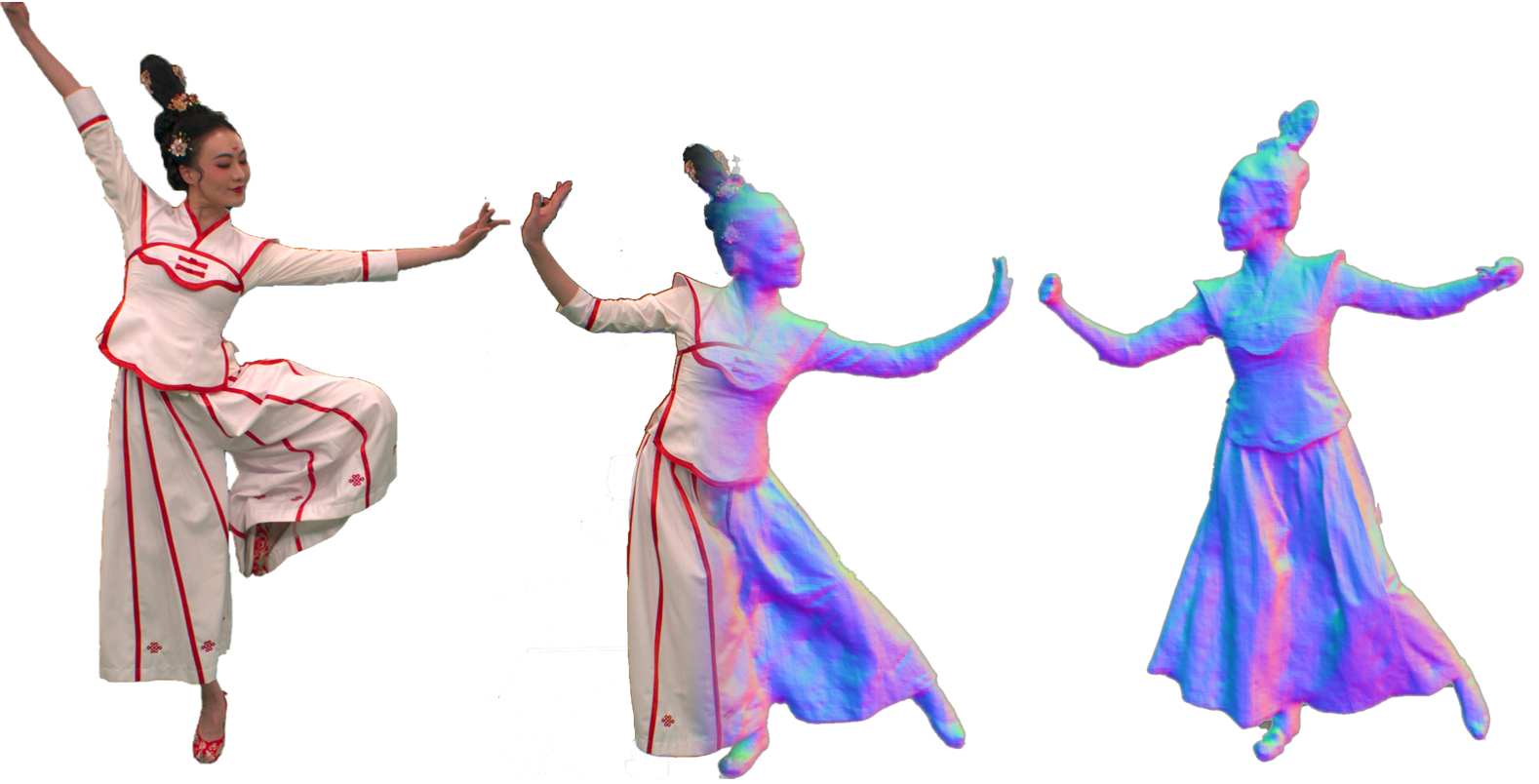}
  \caption{Sampled results from our fast surface reconstruction pipeline. Within minutes training in our PyTorch implementation, our method can generate high-quality geometric reconstruction. }
  \label{fig:neus sample} 
\end{figure}

\paragraph{Human Performance Capture.}
Recently, free-form dynamic reconstruction with real-time performance have been proposed through the traditional non-rigid fusion pipelines. Starting from the pioneering work DynamicFusion~\cite{Newcombe2015} which benefits from the GPU solver to start-of-art multiple camera systems, Fusion4D~\cite{dou2016fusion4d} and motion2fusion ~\cite{motion2fusion}, which extend the non-rigid tracking pipeline to capture dynamic scenes with challenging motions. KillingFusion~\cite{KillingFusion2017cvpr} and SobolevFusion~\cite{slavcheva2018sobolevfusion} proposed more constraints on the motion field to support topology changes and fast inter-frame motions. DoubleFusion ~\cite{DoubleFusion} combines the inner body and outer surface to more robustly capture  human activities by adding body template SMPL~\cite{SMPL:2015} as prior knowledge. Thanks to the human motion representation, UnstructuredFusion ~\cite{UnstructureLan} achieves an unstructured multi-view setup. Function4d~\cite{yu2021function4d} marries the temporal volumetric fusion with the implicit functions to generate complete geometry and can handle topological changes. However, these methods all utilize depth maps that are limited by depth sensor accuracy. Meanwhile, these methods and follow-up work \cite{UnstructureLan, yu2021function4d, jiang2022neuralhofusion} are sensitive to the RGB and the depth sensors calibration.
% that want to perform non-rigid tracking and generate appearances for footprint compression

Based on digit human modeling, some previous works explore supporting free view-point video via mesh compression. The key to geometry-consistent compression is reliably establishing correspondences across all frames stored as a 3D mesh sequence.
In the context of performance capture, the task remains challenging due to large non-rigid deformations that can easily lose track over time. Topology changes and degraded reconstructions (e.g., textureless regions) cause additional problems. Existing shape-only descriptors~\cite{windheuser2014optimal,litman2013learning} are sensitive noise whereas dense shape correspondences assume topology consistency and have been mainly deployed to zero-genus surfaces.
Earlier matching schemes impose strong assumptions including isometric or conformal geometry, geodesic or diffusion constraints, local geometry consistency, etc, under the Markov Random Field (MRF) or Random Decision Forest (RDF) frameworks. Recent neural approaches attempt to train a feature descriptor from multi-view depth maps or even a panoramic depth map to classify different body regions to improve robustness. 

Once constructed, animated meshes can be effectively compressed to further save storage. PCA-based methods~\cite{alexa2000representing,vasa2007coddyac,luo2013compression} aim to identify different geometric clusters of human body (arms, hands, legs, torso, head, etc) whereas~\cite{gupta2002compression, mamou2009tfan} conduct pre-segmentation on the mesh to ensure connectivity consistencies. Spatio-temporal models can further~\cite{mamou2009tfan, ibarria2003dynapack, luo2013compression} predict vertex trajectories for forming vertex groups. A largely overlooked problem though is rendering the sequence: in traditional CG, a single, pre-baked texture with sufficient resolution suffices for rendering all meshes within the sequence. In real captured sequences, however, producing a single texture with comparable quality is nearly impossible due to occlusions, calibration errors of cameras, and color or lighting inconsistencies. Further, using a single texture map loses view dependency. In contrast, our approach recovers human geometry from the multi-view RGB images and perform non-rigid tracking via the traditional pipeline and the neural deformation networks, which can handle human activities naturally. 

\paragraph{Neural Human Rendering.} 
% borrow ideas from fusion but implicitly integrate into networks; history on fusion; how it benefits; why itself is insufficient;
In the photo-realistic novel view synthesis and 3D scene modeling domain, differentiable neural rendering based on various data proxies achieves impressive results and becomes more and more popular.
Various data representations are adopted to obtain better performance and characteristics, such as point-clouds~\cite{Wu_2020_CVPR,aliev2020neural,suo2020neural3d}, voxels~\cite{NeuralVolumes}, texture meshes~\cite{thies2019deferred,liu2019neural,shysheya2019textured} or implicit functions~\cite{kellnhofer2021neural,park2019deepsdf,nerf} and hybrid neural blending~\cite{suo2021neuralhumanfvv,sun2021HOI-FVV, jiang2022neuralfusion}.
More recently, \cite{park2020deformable,pumarola2021d,li2020neural,wang2022fourier} extend neural radiance field~\cite{nerf} into the dynamic setting. \cite{peng2021neural, peng2021animatable,zhao2022humannerf,hu2021hvtr} utilize the human prior SMPL\cite{SMPL:2015} model as an anchor and use linear blend skinning algorithm to warp  the radiance field. Furthermore, ~\cite{sun2021HOI-FVV, jiang2022neuralfusion} extend the dynamic neural rendering and blending into the human-objection interaction scenarios. 
However, for the vast majority of approaches above, dense spatial views still are required for high fidelity novel view rendering.
Image-based blending methods learn blending weight for adjacent views and synthesize photo-realistic novel views in a lightweight way. ~\cite{wang2021ibrnet} learns the blending weight to acquire generalization ability. ~\cite{suo2021neuralhumanfvv} uses the occlusion map as guidance for blending weight estimation. ~\cite{jiang2022neuralfusion} combine the image-based blending with the per-vertex texture to solve the occlusion problem.
Comparably, our neural blending scheme with spatial-temporal information which is obtained by canonical space rendering,  while compressing the memory to the extreme, we can also ensure the fidelity of our blending results.
	\section{Instant Neural Surface Reconstruction}\label{sec:Reconstruction}

\subsection{Preliminary}
Recent neural implicit advances, like NeRF~\cite{nerf} and NeuS~\cite{wang2021neus}, exploit the coordinate-based representation to model the scene by querying points' attributes with their 5D coordinates $(x,y,z,\theta,\phi)$. Such representation significantly improves traditional image-based modeling or rendering in an end-to-end and self-supervised manner.
Specifically, NeuS substitutes the density field in the original NeRF with a signed distance field (SDF), allowing for more deterministic and accurate geometry extraction. To maintain the compatibility with traditional mesh-based workflow, we propose a variant of such SDF-based scheme in our neural surface reconstructor. Recall that, the SDF-based geometry surface $\mathcal{S}$ of the scene is defined by the zero-set of its SDF values:
\begin{equation}
    \mathcal{S} = \{ (x,y,z)\in \mathbb{R}^3|f(x,y,z) = 0 \},
\end{equation}
where $f$ is the signed distance function. 
Similar to previous works, we then supervise the geometry learning according to a volume rendering scheme. 
Specifically, similar to NeRF, we compute a pixel's color $\hat{C}$ via accumulating the geometry and appearance attributes of $n$ sample points $\{\Vec{p}(t) = \Vec{o}+t_i\Vec{d}\mid i=0,1,...,n\}$ along the ray.  
However, in SDF-based representation, the density value is not defined. 
We hence follow NeuS~\cite{wang2021neus} to replace the densities with the opaque density $\rho$ which is based on SDF value $f(\Vec{p}(t))$ at position $\Vec{p}(t)$ and $\Phi_b(x) = 1/ (1+e^{-bx})$ known as the cumulative density distribution, where $b$ is a trainable hyper parameter and gradually increases to a large number as the network training converges. To obtain discrete counterparts of the opacity and weight function, we still need to adopt an approximation scheme, which is similar to the composite trapezoid quadrature. By using opaque density $\rho$, the opacity values $\alpha$ are defined in discrete form by
\begin{equation}
    \alpha_i = \max\left(1 - \frac{\Phi_b(f(\Vec{p}(t_{i+1})))}{\Phi_b(f(\Vec{p}(t_{i})))}, 0\right).
\end{equation}
Thus, the final color of a ray is computed by classical volume rendering equation using the redefined $\alpha$ values,
\begin{equation}
    \hat{C}(t) = \sum_{i=1}^{n}T(t_{i})\alpha(t_{i})c(t_{i})
\end{equation}
where $T_i$ is the discrete accumulated transmittances defined by $T_i =\Pi_{j=1}^{i-1}(1-\alpha_j)$.

\subsection{Truncated SDF Hash Grids}

\begin{figure}[t]
  \includegraphics[width=\linewidth]{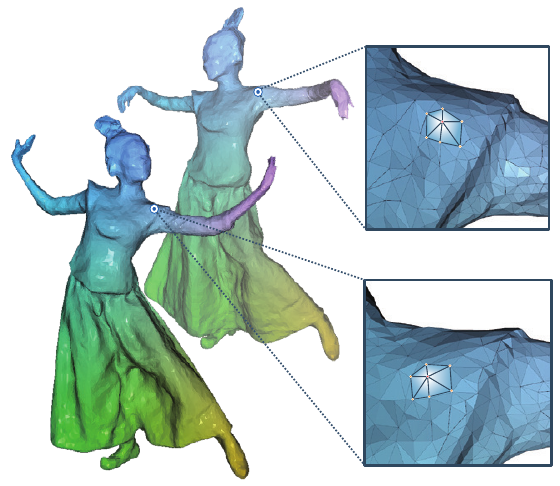}
  \caption{Our neural animation mesh keeps the mesh representation through marrying the traditional non-rigid fusion and the neural rendering advance.}
  \label{fig:am_teaser} 
\end{figure}

Here, we introduce a neural surface reconstructor called Instant-NSR, analogous to Instant-NGP, which can efficiently generate high-quality surfaces from the dense multi-view images. As shown in Fig.~\ref{fig:neus pipeline}, it combines the hash encoding technique~\cite{mueller2022instant} with TSDF (Truncated SDF)-based volumetric rendering similar to the one in NeuS~\cite{wang2021neus}. The use of hash encoding buys us a very efficient representation learning and querying scheme, and the TSDF representation can greatly increase the stability of network training. 

Specifically, given a sample point $\Vec{p}(\Vec{t})$ on a ray $\Vec{t}$, we first acquire feature vectors from the $L$-level hash grids by interpolating the trainable features at the eight vertices of corresponding grids. 
We concatenate acquired feature vectors into $F_{hash}\in \mathbb{R}^{L\times d}$ and then feed it into our SDF network $m_s$ which consists of a shallow MLP. 
$m_s$ output a SDF value $x$ of the sample point, which is formulated as.
\begin{equation}
    (x, F_{geo}) = m_s(p, F_{hash}).
\end{equation}

Naively applying SDF representation to the hash encoding framework will introduce several convergence problems during optimization.
Original SDF-based methods utilize cumulative density distribution $\Phi_b(x) = 1/ (1+e^{-bx})$ to the compute alpha value, which has numerical issues. 
The term $-bx$ will be a large positive number when $b$ is increased, and it leads $e^{-bx}$ close to infinity. 
This kind of numerical instability makes loss function diverge easily during training.

\begin{figure*}[t]
  \includegraphics[width=\linewidth]{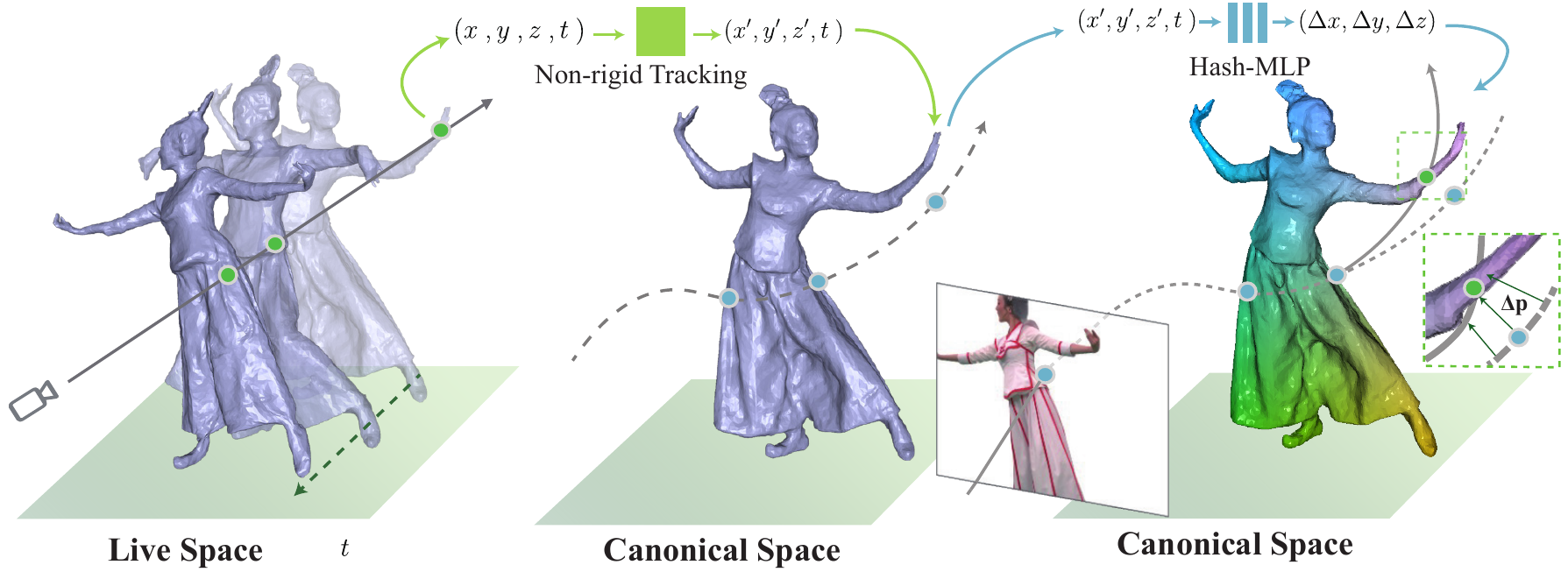}
  \caption{The illustration of backward warping pipeline. We utilize two stages strategy to perform our neural tracking. In coarse stage~\ref{sec:track}, we transform points in live space back to the canonical via the traditional non-rigid-tracking. In fine-tuning stage~\ref{sec:net}, our deform net based on hashing encoding learns the residual displacement.}
  \label{fig:am_pipeline} 
\end{figure*}

To avoid value overflow in $\Phi_b(x)$, we introduce Truncated Signed Distance Function (TSDF) denoted as $\hat{x}$.
Since TSDF has a value range of $-1$ to $1$, this modification ensures numerical stability. 
We apply sigmoid function $\pi(\cdot)$ on network SDF outputs instead of straight-forward truncating SDF to better convergence and avoid vanishing gradient problems.
\begin{equation}
    \pi(x) = \frac{1-e^{-bx}}{1+e^{-bx}}.
\end{equation}
So now $\Phi_b(x) = 1/ (1+e^{-b\cdot\pi(x)})$.

Similar to Instant-NGP~\cite{mueller2022instant}, we feed geometric coding $F_{geo}$, the position $\Vec{p}$ and the view direction $\Vec{v}$ of the sample point into a color network $m_c$ to predict its color $\hat{C}$. 
In addition, we include the normal $\Vec{n}$ of the point as part of the input, which is computed by the gradient of the SDF $\Vec{n}=\nabla \hat{x}$. 
Our appearance prediction is formulated as:
\begin{equation}
    \hat{C} = m_c(\Vec{p}, \Vec{n}, \Vec{v}, F_{geo}).
\end{equation}
The reason we introduce $\Vec{n}$ is to regularize the output SDF implicitly based on an observation that the color network is biased to output similar colors for neighboring sample points if their normals are also close. 
Since we define the normal as the first order derivatives of the SDF, The gradients of normals can be backpropagated to the SDF. 
Adding normals to the input can make the reconstructed surface more even, especially for texture-less areas.

To learn the parameters of the hash table, network $m_s$ and $m_c$, we exploit photometric loss to supervise their training.
We assume the sample number along each ray is $n$ and the batch size is $b$. The loss function is defined as:
\begin{equation}
    \mathcal{L} =  \mathcal{L} _{color} + \lambda\mathcal{L}_{eik},
\end{equation}
where
\begin{equation}
    \begin{aligned}
        \mathcal{L} _{color} = \frac{1}{b}\sum_{i}^{b}\mathcal{R}(\hat{C}, C),\\
        \mathcal{L}_{eik} = \frac{1}{nb}\sum_{k,i}^{n,b}(|\Vec{n}| - 1)^2
    \end{aligned}
\end{equation}
Similar to Instant-NGP implementation, we choose $\mathcal{R}$ as Huber loss in our training process for a more stable training process. To regularize the geometry, We add an Eikonal loss~\cite{gropp2020implicit} on the normal of sampled points.
As shown in Fig.~\ref{fig:neus sample}, our PyTorch implementation obtains high-quality surface result within 10 minutes, favorably compared to existing tools~\cite{AGISOFT} or neural arts~\cite{wang2021neus} in terms of efficacy and efficiency. 
For CUDA-based acceleration similar to Instant-NGP, we adopt finite difference function to approximate the gradient for efficient normal calculation in Eqn.~6 and 8. Such strategy avoids tedious backgrogation of second-order gradients, which is still not fully-supported yet in existing libraries like tiny-CUDANN~\cite{tiny-cuda-nn}. To stimulate future work for more faithful CUDA implementation, we will make our version publicly available.

    \section{Neural Animation Mesh} \label{sec:am}

Even though our Fast Neural Surface Reconstruction (Sec.~\ref{sec:Reconstruction}) provides high-accurate geometry for rendering, saving each frame's geometry individually is inefficient due to its large memory footprint, limiting the application scope. 
To this end, we set out to construct topologically consistent meshes from high-accurate geometry for further data compressing, and also to enable temporal editing effects.
The generation of animation mesh is based on finding geometry correspondences among frames. 
As shown in Fig.~\ref{fig:am_pipeline}, we propose a neural tracking pipeline that marries the non-rigid tracking and neural deformation net in a coarse-to-fine manner. 
It consists of two stages as described below. 

\subsection{Coarse Tracking Stage} \label{sec:track}

\begin{figure}[t]
  \includegraphics[width=\linewidth]{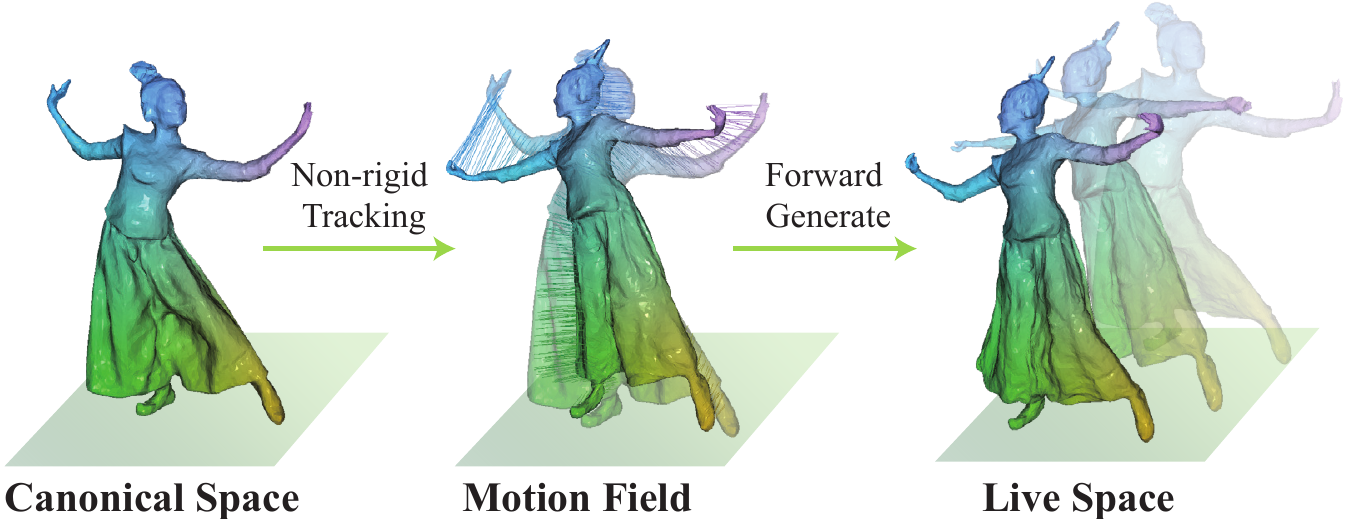}
  \caption{The process of our animation mesh generation. We re-optimize the motion field and perform forward Linear Blend Skinning.}
  \label{fig:am_forward} 
\end{figure}

In the coarse tracking stage, we adopt the traditional non-rigid tracking~\cite{guo2015robust,Newcombe2015,DoubleFusion,FlyFusion,UnstructureLan,jiang2022neuralfusion} based on embedded deformation to establish the coarse correspondences. 
As shown in Fig.~\ref{fig:am_teaser}, we build a canonical space at the first frame and let the first frame's reconstruction be the canonical mesh as the tracking reference.  
Based on the graph-based representation ~\cite{Newcombe2015}, we uniformly sample the embedding deformation (ED) nodes~\cite{sumner2007embedded} by computing the geodesic distance on the canonical mesh and use these nodes to drive it into the target frames.
Meanwhile, we parameterize human motion $H=\{T_i,x_i\}$ with ED nodes, where $x_i$ is the sampled ED nodes coordinates and $T_i$ is the rigid transformations. 
Once obtaining the optimized motion $H$, we can transform the vertices of canonical mesh to live space using the Linear Blend Skinning (LBS)~\cite{SMPL:2015}:
\begin{equation}
    \mathbf{v}_j^{\prime}=\sum_{i\in \mathcal{N}(\mathbf{v}_j)}\omega(\mathbf{x}_i,\mathbf{v}_j)T_i\mathbf{v}_j.
    \label{eq_vj}
\end{equation}
%  w(vi, xj) = max(0, (1-(d(vi, xj)/r)^2)^3)
where $\mathbf{v}_j$ is a vertex on the canonical mesh, $\mathcal{N}(\mathbf{v}_j)$ represents the neighboring ED nodes of the vertex, $\omega(\mathbf{x}_i,\mathbf{v}_j)=\left(1-\left\|\mathbf{v}_{j}-\mathbf{x}_{i}\right\|_{2}^{2} / r^{2}\right)^3$ is the influence weight of the $i$-th node $x_i$ to $v_j$. $r$ is the influence radius and is set to 0.075 in our experiments. 
\paragraph{Warping Field Estimation.}
%When initializing the ED graph $G$ and blending weight $w$, 

However, it is unrealistic to directly compute the motion $H$ between the canonical frame and current frame due to their large motion differences that will lead non-rigid tracking to be failed. 
We instead track adjacent frames and then propagate the optimized local warping fields $W$ to the canonical space to obtain the motion $H$ for each frame.
Following the traditional non-rigid tracking pipeline~\cite{Newcombe2015, li2009robust}, we search point-to-point and point-to-plane correspondences between frame $t-1$ and frame $t$ using non-rigid Iterative Closest Point (ICP) algorithm. 
Since we already have per-frame geometry, we can directly use the geometry to perform tracking instead of using depth inputs. 
The advantages of utilizing the geometry as input have two folds. 
First, it prevents the fusion algorithm from suffering input noises caused by the depth sensor; 
And second, we are able to establish initial local correspondences in Euler space for tracking optimization, instead of finding correspondence only on individual rays~\cite{KinectFusion}.
The energy function for optimizing the warping field $W_t$ is formulated as: 
\begin{equation}
E(W_t,V_{t-1},M_t,G) = \lambda_{\mathrm{data}}E_{\mathrm{data
}}(W_t,V_{t-1},M_t) + \lambda_{\mathrm{reg}} E_{\mathrm{reg}}(W_t,G), 
\end{equation}
where $V_{t-1}$ is the animation mesh warped from the canonical space using the warp field $W_{t-1}$; 
$M_t$ is the reconstructed geometry of current frame obtained from (Sec.~\ref{sec:Reconstruction}). 

The data term is used to minimize the fitting error and formulated as:
\begin{equation}
   E_{\mathrm{data}}=\sum_{\mathbf{v}_{j} \in \mathcal{C}} \lambda_{\text {point }}\left\|\mathbf{v}_{j}^{\prime}-\mathbf{c}_{j}\right\|_{2}^{2}+\lambda_{\text {plane }}\left|\mathbf{n}_{\mathbf{c}_{j}}^{\mathrm{T}}\left(\mathbf{v}_{j}^{\prime}-\mathbf{c}_{j}\right)\right|^{2},
   \label{am_loss}
\end{equation}
where $\mathbf{v}_{j}^{\prime}$ is the warped vertex $\mathbf{v}_{j}$ from canonical space, and 
$\mathbf{c}_{j}$ is the corresponding vertices on the current frame mesh. 
And $\mathcal{C}$ denotes the set of vertices pairs obtained from initial correspondences finding in non-rigid ICP. The weights $\lambda$ balance the relative importance of different terms. In our experiments, we set $\lambda_{\mathrm{data}} = 1$, $\lambda_{\mathrm{reg}} = 20$,  $\lambda_{\mathrm{point}} = 0.2$ and $\lambda_{\mathrm{reg}} = 0.8$.
\\
To constrain the smoothness of the ED node motion, we adopt a locally as-rigid-as-possible regularization:
\begin{equation}
E_{\mathrm{reg}}=\sum_{\mathbf{x}_{i}} \sum_{\mathbf{x}_{j} \in \mathcal{N}\left(\mathbf{x}_{i}\right)} \omega\left(\mathbf{x}_{i}, \mathbf{x}_{j}\right)\psi_{\mathrm{reg}}\left\|\mathbf{T}_{i}\mathbf{x}_{i}-\mathbf{T}_{j}\mathbf{x}_{i}\right\|_{2}^{2},
\end{equation}
where $\mathbf{x}_{i},\mathbf{x}_{j}$ are ED nodes which share the edge on ED graph $G$. $ \omega\left(\mathbf{x}_{i}, \mathbf{x}_{j}\right)$ defines the weight with the edge. $\mathbf{T}_{i},\mathbf{T}_{j}$ are transformations of ED nodes. $ \psi_{reg}$ is the discontinuity preserving Huber penalty.
We then solve the non-linear least squares problem via Gauss-Newton solver on GPU. Please refer to~\cite{Newcombe2015,DoubleFusion} for more details.

\begin{figure}[t]
	\includegraphics[width=\linewidth]{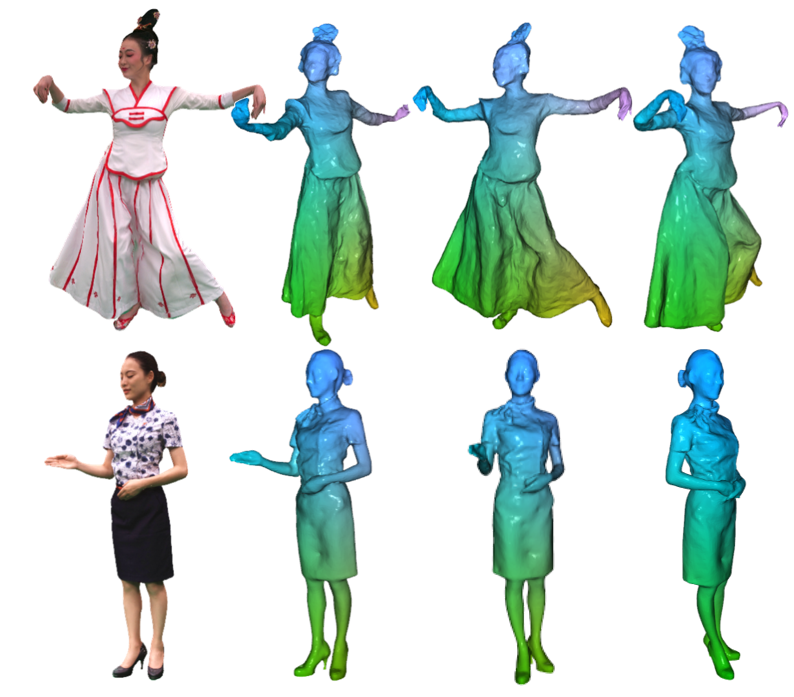}
	\caption{Our neural tracking results. Our pipeline can handle different types of human activities.}
	\label{fig:am_gallery}
\end{figure}

\subsection{Tracking Refinement Stage}  \label{sec:net}

Our explicit coarse non-rigid tracking stage has expressive representation power in terms of complex motion, but still suffers from the limited degrees of freedom.
Therefore, in the refinement stage, we deploy a neural scheme, called deform net, to allow more degrees of freedom for geometry tracking, correct the misalignment and preserve fine-grained geometric details, which is significant progress from the existing work~\cite{Newcombe2015,UnstructureLan,KillingFusion2017cvpr,yu2018doublefusion}.

We use two steps to achieve this goal. 
First, we leverage the original Instant-NGP~\cite{mueller2022instant} to obtain a radiance field $\phi^o$ in the canonical space in a rapid way: $(\mathbf{c}, \sigma)=\phi^o(\Vec{p},\Vec{d})$, where $\mathbf{c}$ denotes the color field and $\sigma$ denotes the density field, $\Vec{p}$ is a point the canonical space and $\mathbf{d}$ is the ray direction. 
After obtaining this radiance field, we froze the model parameters. 
Second, we train a deformation network $\phi^d$ to predict per-frame warping displacement and further finetune the motion $H$ of a frame. 
At the core of the second step is to build a connection between the coarse motion estimation and photometric loss scheme so that we can optimize the warping displacement using input images in end-to-end supervision. 
We introduce a differentiable volume rendering pipeline cooperated with tracked ED nodes to realize the supervision.

\begin{figure*}[ht]
  \includegraphics[width=\linewidth]{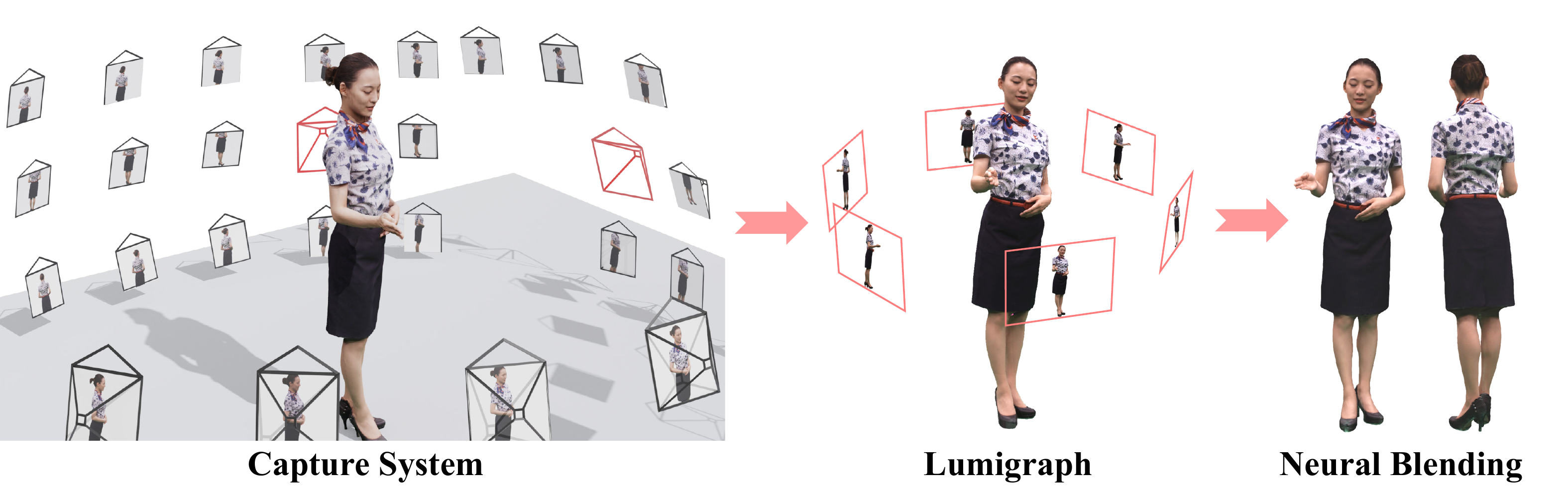}
  \caption{The pipeline of neural texture blending. The raw images obtained from black cameras in the capture system are fed into lumigraph rendering system. We pick six novel views (red camera in the capture system) as the output views of lumigraph. These output views, together with the corresponding depth images, are sent into the neural blending module to generate novel view images. }
  \label{fig:lumigraph} 
\end{figure*}

\begin{figure}[t]
  \includegraphics[width=\linewidth]{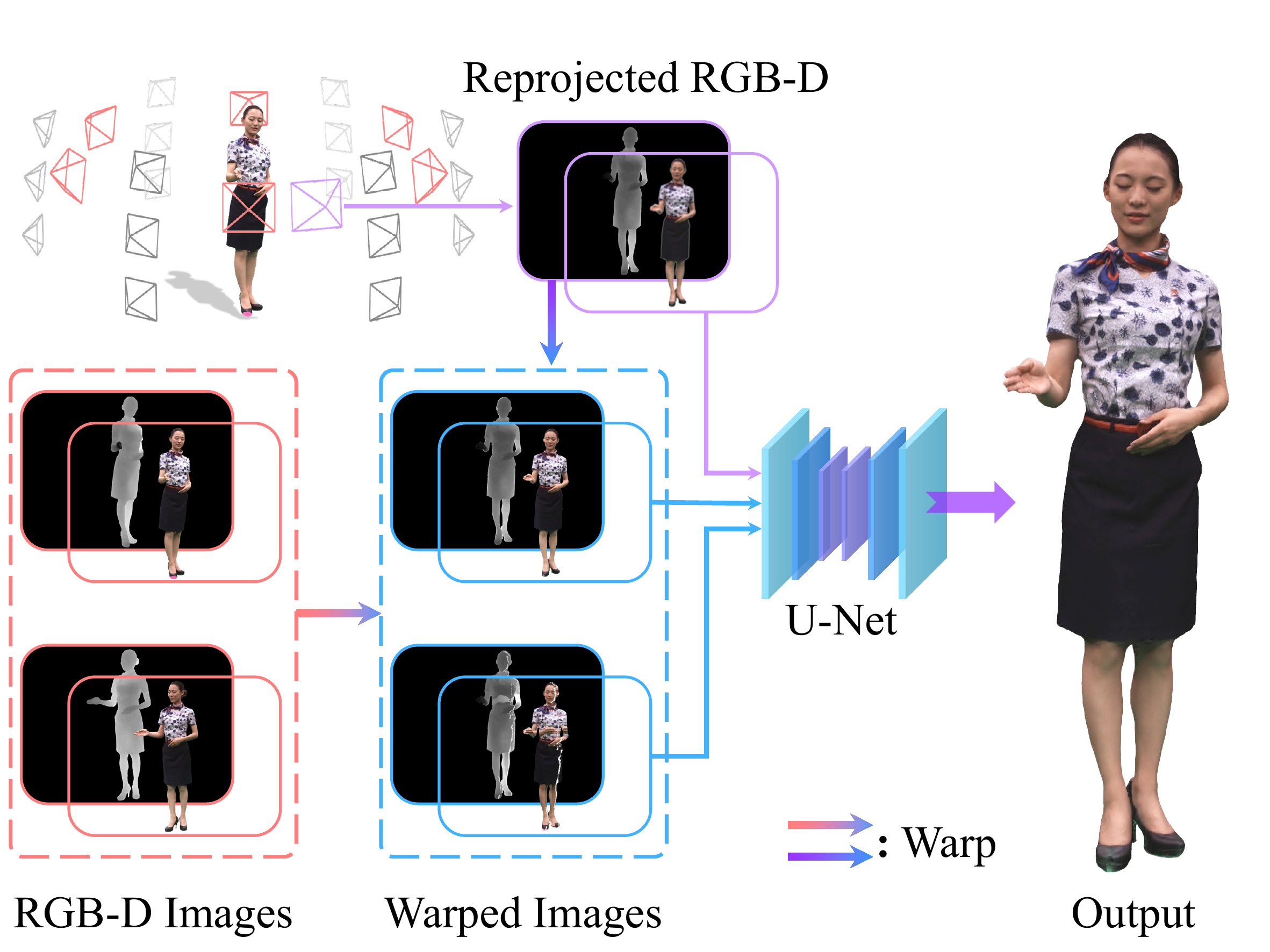}
  \caption{The detail of the neural blending module. Once the novel view (purple) is chosen, a coarse RGB image and precise depth can be obtained from textured mesh. The depth warps with the RGB-depth images from two of six neighbor views (red) and generates the warped images. The U-net takes as input the coarse RGB and the warped images and outputs the novel view image. }
  \label{fig:warp} 
\end{figure} 

Since having a coarse motion $H$ , we transform ED nodes from canonical frame to current frame and sample points along with pixel rays of the target view.  
For each sample point $\Vec{p}$, we find its $k$-nearest ED nodes and interpolate their inverse warping to warp $\Vec{p}$ back to canonical space. 
The warped point in canonical space is denoted as $\Vec{p}'$. 
We take $\Vec{p}'$ and the timestamp of current frame $t$ as the input of the deformation network $\phi^d$. 
The network $\phi^d$ will predict a displacement $\Delta \Vec{p}'$. 
%, which is formulated as $\Delta \Vec{p}' = \phi^d(\Vec{p}', t)$. 
We adopt the hash encoding scheme~\cite{mueller2022instant} to $\phi^d$ and make its training and inference efficient. 
The density and color of the sample point $\Vec{p}$ is then formulated as: 
\begin{equation}
   (\mathbf{c}, \sigma)=\phi^o(\Vec{p}' + \phi^d(\Vec{p}', t), \Vec{d}). 
\end{equation}
Fig.~\ref{fig:am_pipeline} demonstrate our refinement pipeline. 
This pipeline allows us to accumulate colors of pixel rays and apply the photometric loss to supervise the network training.    

\begin{figure*}[thp]
	\includegraphics[width=\linewidth]{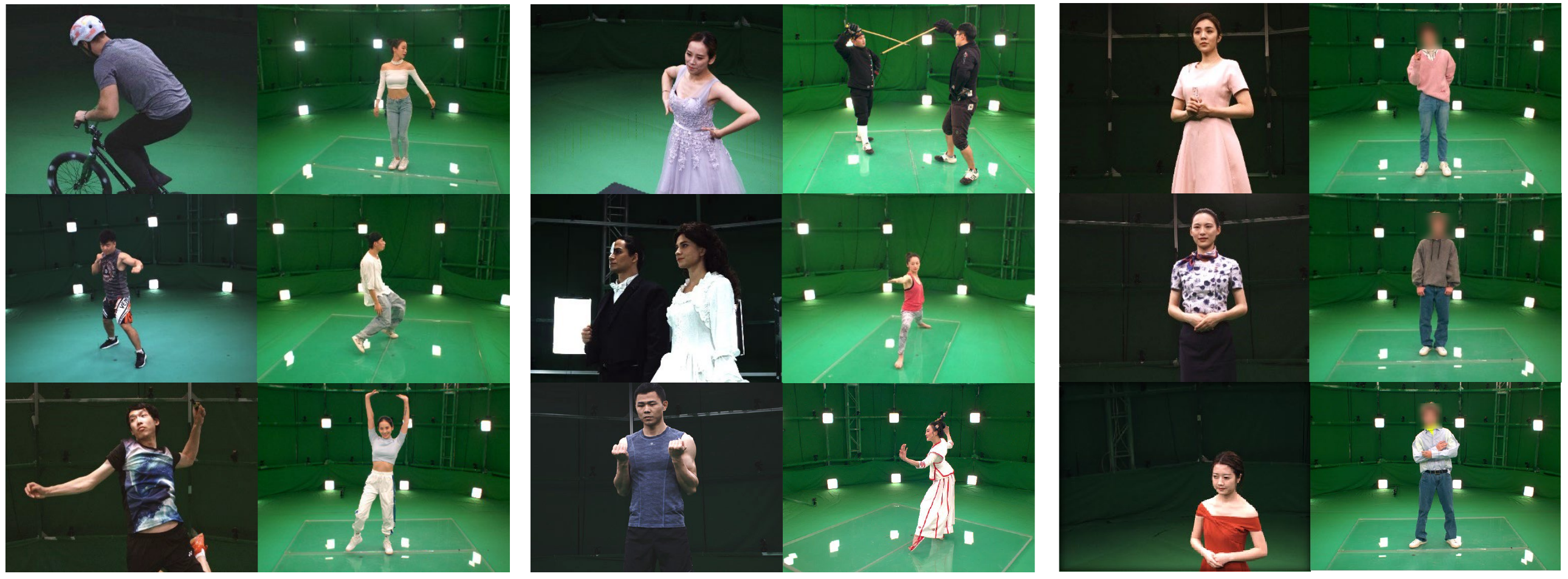}
	\caption{Data examples were captured by our multi-view greenscreen dome with 80 synchronized high-resolution RGB cameras. Our dataset includes a variety of human performances under various motions, garments, and interactions.}
	\label{fig:data_gellery}
\end{figure*}

\subsection{Animation Mesh Generation}

The use of a deformation network provides us with better correspondences from the current frame to the canonical frame. 
We can exploit the fine-grained correspondences as better priors in non-rigid tracking algorithm to re-optimize the motion as $H'$. 
To generate the animation mesh for the current frame, we just need to apply $H'$ on the vertices of the canonical mesh and remain the topological relation to be the same.  

As shown in Fig.~\ref{fig:am_forward}, we utilize the refined warp field parameters to track and acquire the live frame animation mesh. Instead of saving the original mesh sequences and network parameters, we only need to save the canonical mesh and motion field which can support playback and real-time rendering and editing.
\\

\begin{figure*}[thp]
	\includegraphics[width=0.95\linewidth]{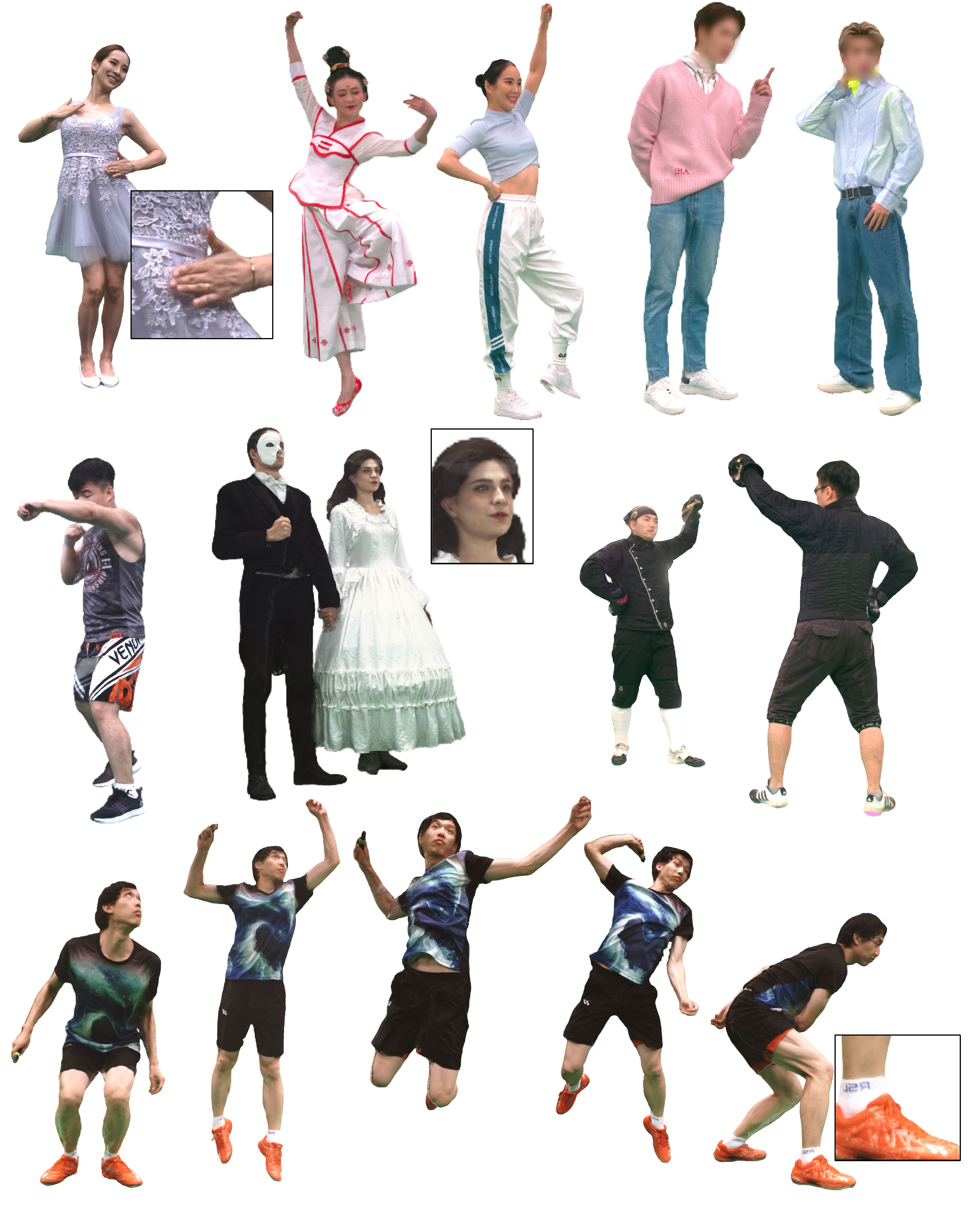}
	\caption{Gallery of our example results. Our neural pipeline enables efficient and photo-realistic rendering of  human performances and is compatible with conventional mesh-based pipeline. Note that the results are all rendered at novel viewpoints.}
	\label{fig:result_gallery}
\end{figure*}

\begin{figure*}[thp]
	\includegraphics[width=\linewidth]{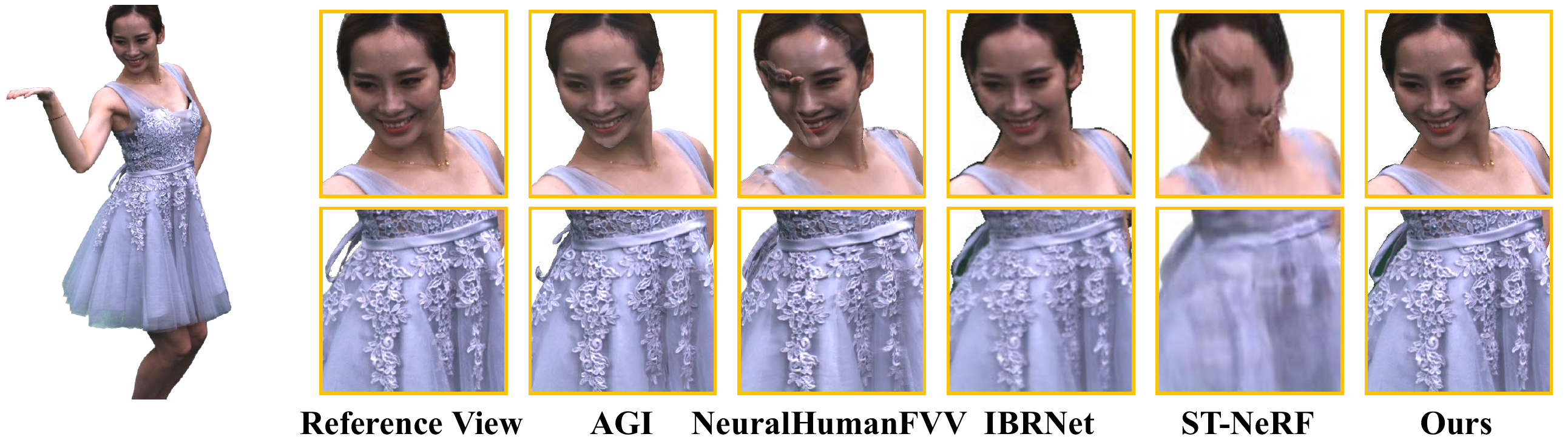}
	\caption{Qualitative Comparison against different rendering methods, including AGI, NeuralHumanFvv, IBRNet, and ST-NeRF. Note that our approach achieves more realistic rendering with sharper textured details.}
	\label{fig:comparison blending}
\end{figure*}

    \section{Rendering} \label{sec:Rendering}

The accurate geometry from neural animation mesh reconstruction enable high-quality rendering for various rendering approaches, such as textured mesh rendering and Image-based rendering (IBR). 
These methods can produce plausible image quality in novel views based on our neural animation mesh but suffer from either the lack of view-dependent effects or a large storage of reference views. 
In this section, We propose a method combining explicit 2D texture map and implicit neural texture blending which is streamable while retaining high rendering quality. 
To this end, our proposed method reduces the bit rates of the 4D video sequence to enable streamable applications and aims to have lower storage.

\subsection{Appearance Distillation}
Recall that our neural animation mesh can reduce the storage of geometry, but we still have dozens of reference views whose videos cost storage space. 
The key idea is to distillate the appearance information for all reference views into fewer images. 
We achieve this goal via image based rendering. 
Specifically, we pre-define six virtual viewpoints evenly around and towards the performer in the center.
These views have sufficient field of view so that the entire performer can been seen in the views as shown in Fig.~\ref{fig:lumigraph}. 
We synthesis these view images using unstructured lumigraph~\cite{gortler1996lumigraph} that calculates pixel blending from neighbor reference views based on a given geometry. 
Synthesised views code the appearance information of original reference views via blending pixel from different body parts and other views. 
The number of views that need to be stored is reduced to six in this way. 
However, this kind of reduction causes the information loss that will degrade the quality of the results during distillation. 
So we deploy a neural blending method to compensate the information loss via neural networks. 

\subsection{Neural Blending}
We introduce a neural blending scheme that can blend fine-grained texture details from adjacent virtual views into the novel view. 
Similar to \cite{suo2021neuralhumanfvv}, the input of neural blending not only includes two adjacent views of the target camera but also contains their depth maps for occlusion awareness. 
However, some self-occluded parts may not be observed by any adjacent virtual view due to the view sparsity. 
To fill this kind of area in the novel view, we also adopt the textured animation mesh rendering results as part of the blending component. 
Fig.~\ref{fig:warp} demonstrates the pipeline of our blending module. 

\paragraph{Occlusion Awareness} Most of the appearance information in a target view can be observed by its two adjacent virtual views and the missing part due to self-occlusion can be recovered using 2D texture map in our sparse view re-rendering setting.
Based on this assumption, we propose a hybrid method by using selected views to re-render animated mesh through a blending network.
Given $k$ input images around the performer, we first generate the depth maps of the target view ($Dr_t$), the two input views ($Dr_1$ and $Dr2$, respectively), and coarse rendering images at target view $I_r$ using the textured mesh described in Sec.~\ref{sec:am}.
Then, we use $Dr_t$ to warp the input images $I_1$ and $I_2$ into the target view, denoted by $I_{1,t}$ and $I_{2,t}$ . 
We also warp source view depth maps into target view and obtain $ Dr_{1,t}$ and $Dr_{2,t}$ so as to obtain the occlusion map $O_i = Dr_t-Dr_{i,t} (i = 1, 2),$ which implies the occlusion information.

\begin{table}[t]
	\centering
	\caption{Breakdown of processing per-frame time in each stage, averaged over a whole captured sequence.}
	\begin{tabular}{lcc}
		\toprule
		Stage & Action & Avg Time\\
		\midrule
		Preprocessing &  background matting& $\sim$ 1 min \\ \hline
		Neural Surface &   fast surface reconstruction& $\sim$  10 mins \\ \hline
		\multirow{2}{*}{Neural Tracking} &  coarse tracking & $\sim$  57 ms \\
		& neural deformation& $\sim$  2 mins \\ \hline
		\multirow{3}{*}{Neural Rendering} &  2D texture generation & $\sim$ 30s \\
		& sparse view generation & $\sim$ 2s \\
		& neural texture blending & $\sim$  25ms \\
		\bottomrule
	\end{tabular}
	\rule{0pt}{0.01pt}
	\vspace{-0.3cm}
	\label{breakdown}
\end{table}

\paragraph{Blending Network.} $I_{1,t}$ and $I_{2,t}$ may be incorrect due to self-occlusion and inaccurate geometry proxy. Simply blending them with rendering from 2D texture will raise strong artifacts.
Thus, we introduce a blending network $U_b$, which utilizes the inherent global information from our multi-view setting, and fuse local fine-detailed geometry and texture information of the adjacent input views with the pixel-wise blending map $W$, which can be formulated as:
\begin{equation}
    W = U_b(I_{1,t}, O_1, I_{2,t}, O_2).
\end{equation}
The network structure of $U_b$ is similar to U-Net~\cite{ronneberger2015u} and outputs two channels feature maps $W=(W_1, W_2)$ representing the blending weights of warped images respectively.

For real-time performance, depth maps are generated at low resolution $(512\times512)$. 
Aiming to photo-realistic rendering, we need to upsample both the depth map and blending map to 2K resolution.
However, naive upsampling will cause severe zigzag effects near the boundary due to depth inference ambiguity.
Thus, we propose a boundary-aware scheme to refine the human boundary area on the depth map. Specifically, we use bilinear interpolation to upsample $Dr_t$. 
Then an erosion operation is applied to extract the boundary area.
Depth values inside boundary area are recalculated by using the pipeline as described in Sec.~\ref{sec:am} and form $\hat{Dr}_t$ at 2K resolution. 
Then we warp the original high-resolution input images into the target view with $\hat{Dr}_t$ to obtain $\hat{I}_{i,t}$. To this end, our final texture blending result is formulated as:
\begin{equation}
    I = \hat{W}_1 \cdot \hat{I}_{1,t} + \hat{W}_2 \cdot \hat{I}_{2,t} + (1 - \hat{W}_2 - \hat{W}_2) \cdot I_r, 
\end{equation}
where $\hat{W}$ is the high resolution blending map upsampled by bilinear interpolation directly; $ I_r$ is the image from the corresponding textured animation mesh. 
We generate training samples using Twindom~\cite{twindom} dataset. 
The synthetic training data contains $1,500$ human textured models, and we render 360 novel views around each of them with depth maps. 
During training, we randomly select six views whose relative poses meet the pre-defined relationship as reference views while the others are the ground truth of target views. 
We apply MSE loss between blended image $I$ and the corresponding ground truth $I'$ to supervide the network training. 
	
\externaldocument{method_tracking}

\section{Results}\label{sec:results}

\begin{table}[t]
	\centering
	\caption{\textbf{Quantitative comparison against several rendering methods in terms of rendering accuracy and training time.} Compared with AGI, NeuralHumanFVV, IBRNet,and ST-NeRF, our approach achieves the best performance in \textbf{PSNR},\textbf{SSIM}, \textbf{MAE} and \textbf{LIPIS} metrics. Besides, our method only requires the minutes training time.}
	\begin{tabular}{l|c|c|c|c}
	\multicolumn{4}{c} {}\\
        Method &  PSNR$\uparrow$ & SSIM$\uparrow$   &MAE$\downarrow$    &LPIPS$\downarrow$ \\ \hline
		AGI             & 31.12 & 0.9671 & 1.089 & 0.0464 \\
		NeuralHumanFVV  & 30.42 & 0.9708 & 2.089 & 0.0451\\
		IBRNet          & 31.60 & 0.986 & 1.212 & 0.0380\\
		ST-NeRF         & 27.12 & 0.9561 & 2.153 & 0.0877\\ \hline
		Ours            & \textbf{35.06 }& \textbf{0.997} & \textbf{0.135} & \textbf{0.0053}\\    \hline
    \end{tabular}
\rule{0pt}{0.05pt}

\label{table:comparison rendering}
\vspace{-0.3cm}
\end{table}

In this section, we demonstrate the capability of our approach in a variety of scenarios. 
We first present the collected multi-view dataset and the gallery of our high-quality rendering results. 
We further provide the comparison with previous state-of-the-art methods as well as the evaluation of our main technical components, both qualitatively and quantitatively, followed by the analysis of our results with various immersive free-view experiences. 
The limitations regarding our approach are provided in the last subsection.

\paragraph{Implementation Details and Running-time analysis.} 
Our whole systems and experiments run on onr single NVidia GeForce RTX3090 GPU.
The AR device we used is the latest iPad Pro and we demonstrate our VR rendering using Oculus Quest 2.
We analyze the relative per-frame processing cost of each stage which is averaged in a whole captured sequence.
As shown in Tab.~\ref{breakdown}, neural surface reconstruction is the most time-consuming process which requires 10 minutes per-frame in our Pytorch implementation. 
However, this processing time will be reduced to about 40 seconds in our final CUDA implementation as well as the neural deformation in our neural tracking stage which can be done in 20 seconds.
Our 2D texture generation and sparse view generation can be also seen as a preprocessing operation after outputting animated mesh using our surface reconstruction and tracking algorithm.
At the running time, we can render a 2048$\times$1536 image in real-time at about 10 fps.

\paragraph{Datasets}
Fig.~\ref{fig:data_gellery} provides the gallery of data examples captured by our multi-view greenscreen dome, covering various garments, motions and interactions. 
For each performance, we provide the video sequences captured by 80 pre-calibrated and synchronized RGB cameras at $2048\times1536$ resolution and 25-30 frame-per-seconds. 
The cameras are uniformly arranged around the performer at 4 circles at different latitudes.
We also provide the foreground segmentation of the dynamic performances using an off-the-shelf background matting approach~\cite{sengupta2020background}. 
Fig.~\ref{fig:result_gallery} illustrates several representative rendering results by our pipeline. Note that our approach generates realistic results for the performers' facial expressions and textured garment details like %the tartan suit or 
the skirt with rich lace patterns. Our approach also faithfully reconstructs various challenging motions like sports, jumping, K-pop or dancing with loose clothes, illustrating the robustness and effectiveness of our pipeline.

\subsection{Comparisons}

\begin{figure}[t]
 	\includegraphics[width=\linewidth]{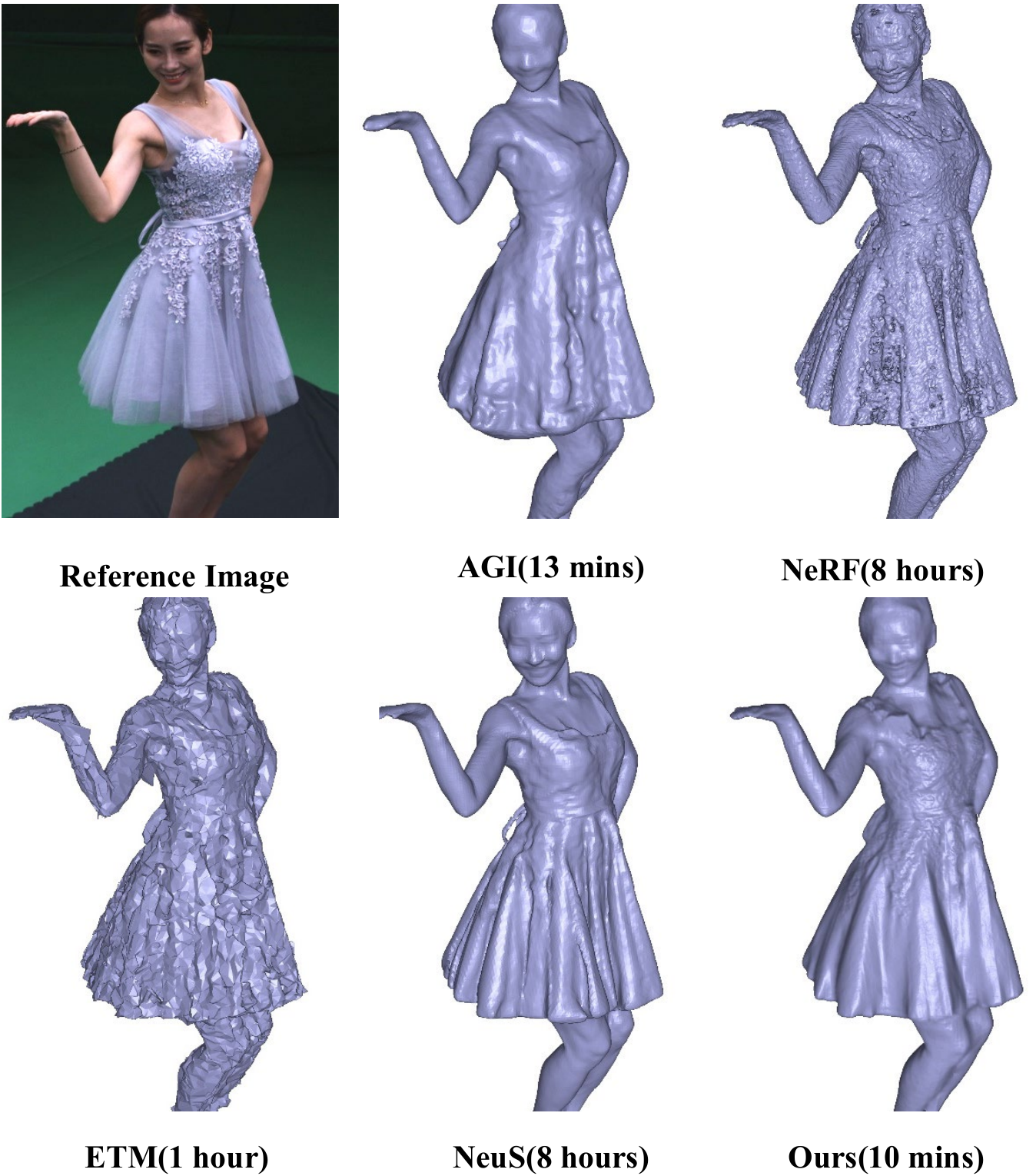}
	\caption{Qualitative comparisons on the geometric reconstruction. Compared to AGI, NeRF, NeuS, and extract triangular 3D models, our method only take 10 minutes training and achieves the superior or comparable geometric reconstruction results.}
	\label{fig:comparison surface}
	\vspace{-0.3cm}
\end{figure}

\paragraph{Rendering comparison}
To demonstrate the overall performance of our rendering pipeline, we compare to the traditional mesh-based modeling pipeline using the commercial software Agisoft PhotoScan~\cite{AGISOFT} and the existing novel view synthesis methods based on neural rendering, including neural image based rendering method IBRNet~\cite{wang2021ibrnet}, NeuralHumanFVV~\cite{suo2021neuralhumanfvv} and dynamic neural radiance field ST-NeRF~\cite{zhang2021editable}.
For a fair comparison, NeuralHumanFVV, IBRNet, and ST-NeRF share the same training dataset as our approach, and we reconstruct the scene to obtain a textured mesh for every single frame in AGI from all the input viewpoints.
For quantitative comparisons, we adopt the peak signal-to-noise ratio (\textbf{PSNR}), structural similarity index (\textbf{SSIM}), mean absolute error (\textbf{MAE}), and Learned Perceptual Image Patch Similarity (\textbf{LPIPS})~\cite{zhang2018unreasonable} as evaluation metrics for our rendering accuracy.
Note that we calculate all quantitative results in the reference view of all captures.
As shown in the Tab.~\ref{table:comparison rendering}, our approach outperforms all the other methods in terms of PSNR, SSIM, MAE and LPIPS, showing the effectiveness of our model to provide a realistic rendering of the complicated dynamic scenes.
For qualitative comparison, we show the novel view rendering results and the nearest input view in Fig.~\ref{fig:comparison blending}.
AGI generates sharper rendering appearance results but is limited by the reconstruction accuracy, leading to severe artifacts in those missing regions, especially near the boundary.
NeuralHumanFVV can provide reasonable results, but it suffers from severe stitching gaps.
ST-NeRF suffers from uncanny blur results due to the  network capabilities in ST-NeRF.
%IBR PROBLEM
IBRNet shows more details in rendering results, but there are also some artifacts due to wrong depth estimation.
In contrast, our approach achieves the most vivid rendering result in terms of photo-realism and sharpness.  

\begin{figure}[t]
	\includegraphics[width=\linewidth]{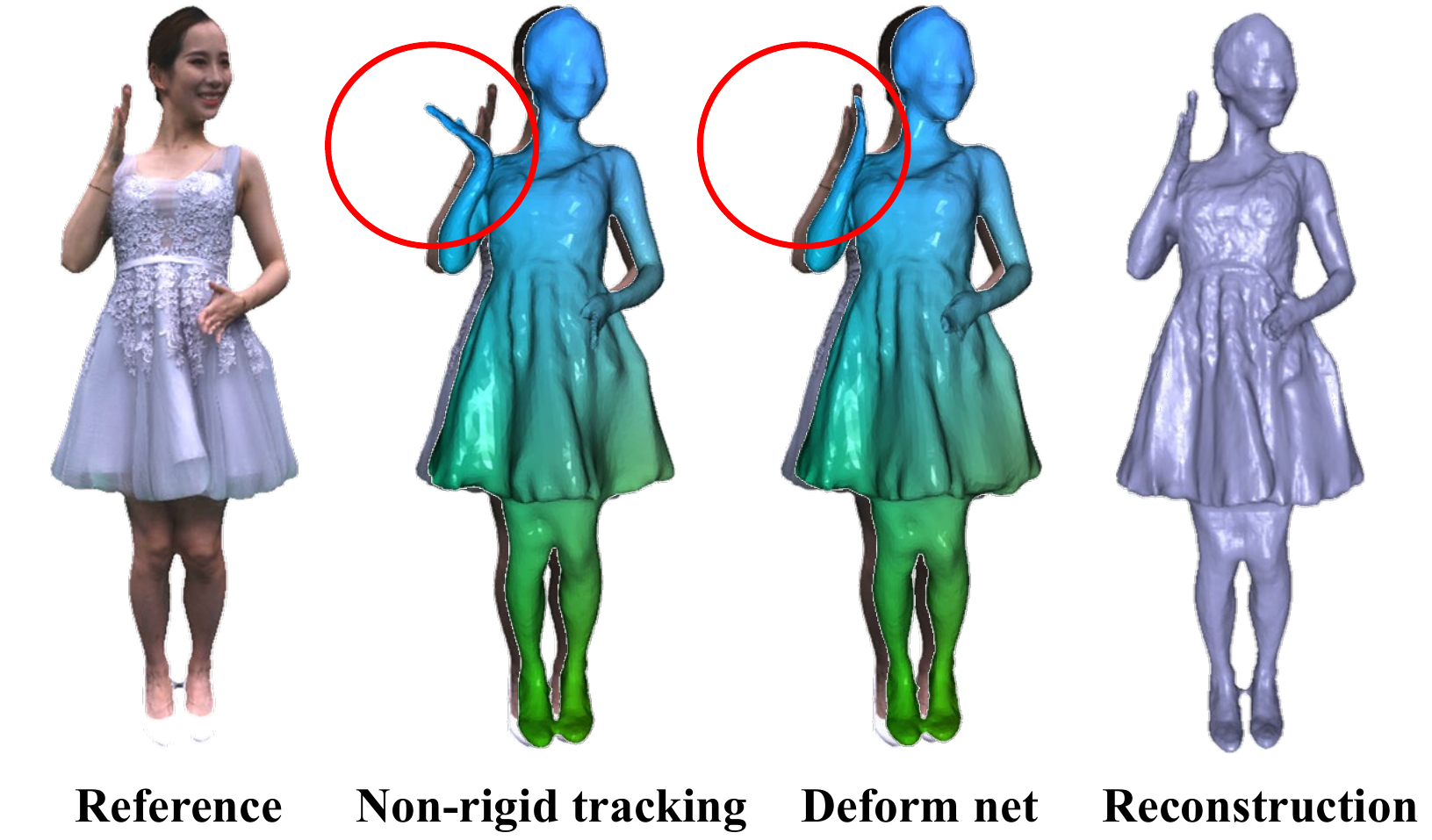}
	\caption{Qualitative evaluation of neural tracking. Compared to coarse non-rigid-tracking, our deformation net aid neural tracking achieves the high accuracy.}
	\label{fig:am_compare}
\end{figure}

\begin{table}[t]
	\centering
	\caption{\textbf{Quantitative comparison against several methods in terms of geometric accuracy.} We use a synthetic data to render multiview images and take the synthetic geometry as the ground truth. Compared with AGI, NeRF, ETM (Extracting Triangular 3D Models), and NeuS, our approach achieves the best balance in accuracy and speed.}
	\begin{tabular}{l|c|c|c|c|c}
		\multicolumn{5}{c} {}\\
		Method &  AGI & NeRF   & ETM    &NeuS &Ours\\ \hline
		%Distance error  & 27.07  & 0.9828  & 0.0053  & 0.0410 & 0.0410 \\  \hline
		Chamfer Distance $\downarrow$  & 0.0421  & 0.0425  & 0.0696  & 0.0410 & 0.0413 \\  \hline
	\end{tabular}
	\rule{0pt}{0.05pt}
	\label{table:comparison geo}
	\vspace{-0.3cm}
\end{table}

\begin{figure*}[thp]
	\includegraphics[width=\linewidth]{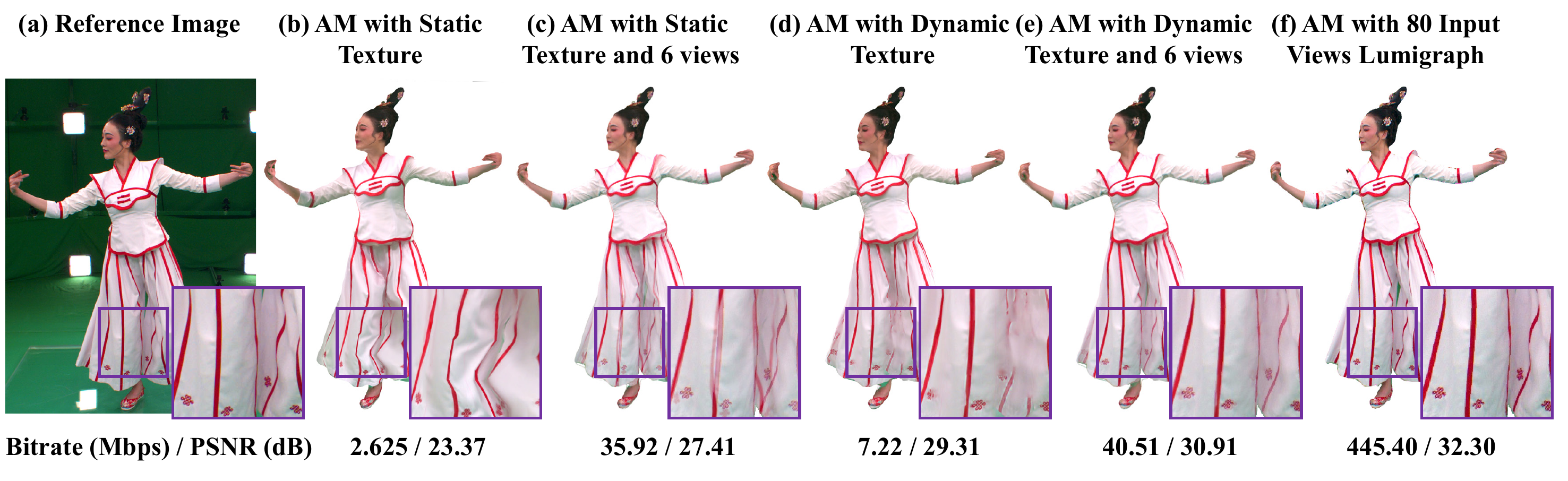}
	\caption{Ablation study. Under a novel view, we compare our animated mesh (AM) on different texture schemes, including static texture map, 6 input views with static texture map, dynamic texture map, 6 input views with dynamic texture map, and 80 input views with Lumigraph. Bitrate measures the amount of data required for rendering per second. These data include compressed meshes (all), compressed textures (d, e) and photos (c, e, f) using H.265 technique or uncompressed texture (b, d). }
	\label{fig:AM and Texture}
	\vspace{-10pt}
\end{figure*}

\paragraph{Geometric comparison}
We then evaluate our fast surface reconstruction pipeline. We compare our approach to the several existing geometry reconstruction methods, including traditional method, such as Agisoft PhotoScan~\cite{AGISOFT}, the implicit representation NeRF~\cite{nerf} based on neural radiance field and NeuS~\cite{wang2021neus} based on neural implicit surfaces, and the mesh based optimization method~\cite{munkberg2021nvdiffrec}.
We perform our comparisons both on qualitative results shown in Fig.~\ref{fig:comparison surface} and quantitative results shown in Tab.~\ref{table:comparison geo}.
As shown in Fig.~\ref{fig:comparison surface}, Agisoft PhotoScan~\cite{AGISOFT} shows the limited performance for reconstructing the silk-like skirt and extracted mesh
from NeRF~\cite{nerf} is noisy since the volume density field has not sufficient constraint on the surface of 3D geometry. 
The extracted mesh from NeuS~\cite{wang2021neus} are closest to our result but it requires about 8 hours training time.
However, our method only needs about 10 minutes training time which is 48x faster than NeuS~\cite{wang2021neus} and achieve equal or even superior results.
To demonstrate the geometric effect of our approach, We also conduct the qualitative comparisons on the synthetic data shown in Tab.~\ref{table:comparison geo} which is measured with the Chamfer distances in the same way as NeuS~\cite{wang2021neus} and UNISURF~\cite{oechsle2021unisurf} and our approach achieve the best balance between geometry accuracy and speed.

\subsection{Ablation Study}

In this subsection, we evaluate the performance of our approach by comparing different designs in our neural tracking and texture blending scheme in quality and quantity.
\paragraph{Neural Tracking.} As shown in Fig. ~\ref{fig:am_compare}, we evaluate our neural tracking pipeline with the same fast reconstruction proxy. We can see during the coarse non-rigid tracking stage, our tracking results are still slightly deviated from the reference view. This effect mainly occurs in places with large displacements such as limbs, which have less ED nodes and can not find the correct correspondences. Through the deformation net and the re-optimize, we can correct this mismatch and the tracking results align with the original reconstruction results.

\paragraph{Texture Blending Scheme.} Here, we evaluate the performance and steaming bit rates of different texture blending schemes. 
As shown in Fig.~\ref{fig:AM and Texture}, single-texture animated meshes have produced reasonable rendering results with our improved tracking algorithm, but still cannot handle subtleties such as wrinkles in clothing and facial expressions, which are difficult to represent with a single texture map.
Animated meshes with dynamic texture can generate sharp rendering results and model facial expressions change, but are still sensitive to the accuracy of geometric reconstruction, especially when geometric details are hard to recover even using neural tracking.
To alleviate the sensitivity of geometric accuracy and make our rendering pipeline more robust while keeping the ability for streaming application, we propose a combination of 2D texture and live time sparse view observation, such as animated mesh with single texture and six selected views and animated mesh with dynamic texture and six selected views.
Animated meshes with dynamic texture and six views input can produce more accuracy rendering results than only using dynamic texture and can introduce view-dependent effects which can not be represented by 2D texture map, 
Animated meshes with lumigraph using 80 views can produce most photo-realistic rendering results but require too much storage.

To further analyze rendering scheme in our approach, we compare the transmission bit rates when applying to steaming task.
Our hybrid rendering pipeline using neural texture blending and explicit 2D texture is a well balance between rendering accuracy and transmission bandwidth.

\begin{figure*}[t]
	\includegraphics[width=\linewidth]{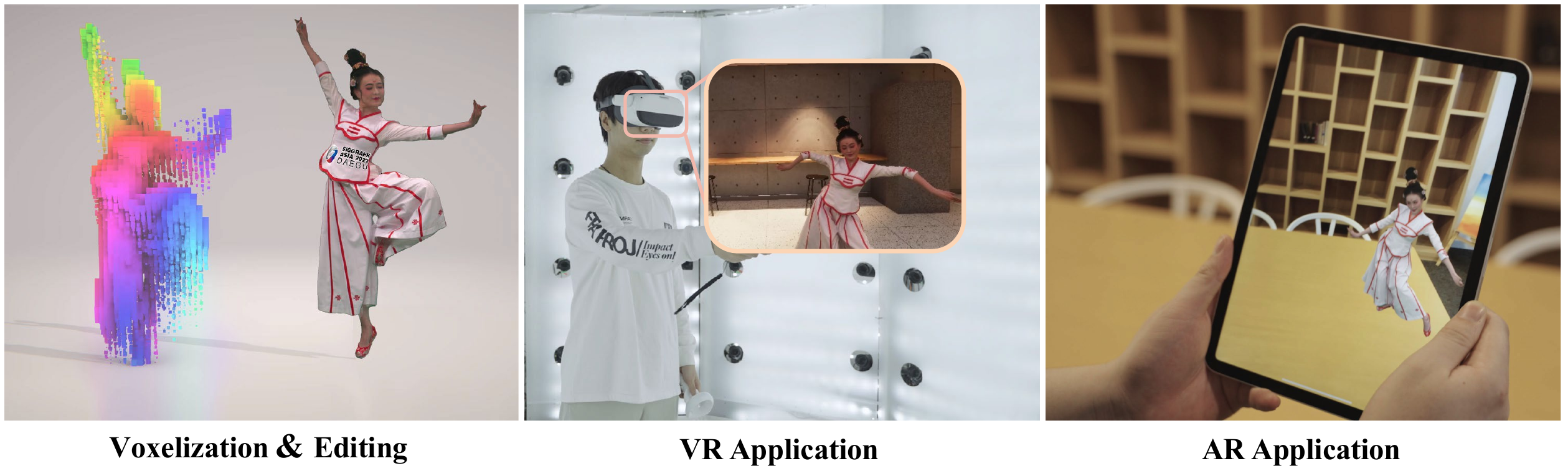}
	\caption{Our approach supports convenient and photo-real post-editing for various fancy visual effects, and enables various immersive experiences of 4D human performance playback on consumer-level VR/AR platforms.}
	\label{fig:Application1}
\end{figure*}
\begin{figure}[t]
	\includegraphics[width=\linewidth]{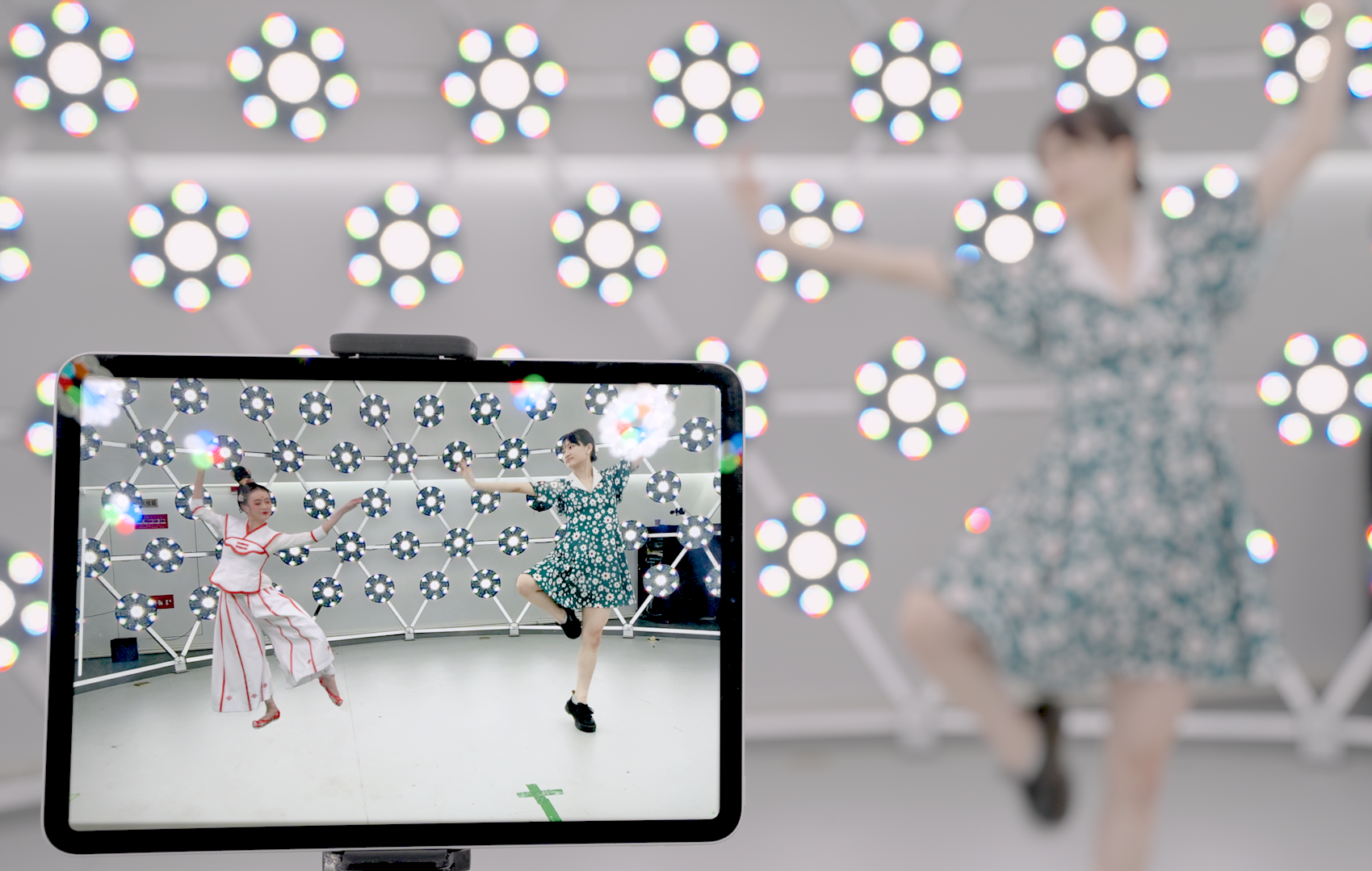}
	\caption{Virtual-real interactions using our approach, where a girl can creatively practice dancing with her favorite actress in an immersive way.}
	\label{fig:AR}
\end{figure}

\subsection{Immersive Experiences} 
Here we showcase various mesh-based applications and photo-real immersive experience with 4D human performance playback in VR/AR using our neural pipeline.

\paragraph{Mesh based Editing.}  
Our generated animated meshes provide rich temporal information and can be readily integrated into mesh-based editing tools for fancy visual effects of 4D human performance. As shown in Fig.~\ref{fig:Application1}, for the dancing sequence, we voxelize the animated meshes of the actress and encode various regions and motions with distinct colors, providing a coherent voxel art effect in the 4D space. Since we have obtained the temporal corresponding information, we can easily draw a logo onto her garments by editing the corresponding texture maps in the canonical UV space. Then, we further combine the edited results with our neural blending process by modifying the corresponding blending maps for the edited region in Eqn.~15. Such strategy enables more natural and photo-real editing in Fig.~\ref{fig:Application1}, which is rare in other implicit neural rendering methods~\cite{NeuralVolumes,nerf}.

\paragraph{Interactive Exploration in VR/AR}
Our light-weight neural blending design enables photo-real human performance rendering on VR/AR platforms. Besides, as a neural extension of meshes-based workflow, our approach naturally support arranging the neural rendered results with traditional CGI background scenes. In Fig.~\ref{fig:Application1}, we demonstrate the application to immersively watch a high-quality talent shows with the VR headset Oculus Quest 2. The viewer can enjoy the shows of the singer and dancer in a photo-real and immersive manner, while in reality the two artists from various places around the world are composited into the same virtual space.

We further demonstrate the interactive experience of 4D performances using ARkit on an iPad Pro. As shown in Fig.~\ref{fig:Application1}, the user can project the dancing actress into the real environment to immersively enjoy the show, and change the views and positions of her performance through fingertips. Moreover, Fig.~\ref{fig:AR} demonstrates the concept of breaking the boundaries between virtual and real worlds, where a girl can creatively practice dancing with her favorite actress in an immersive way using our approach.

\subsection{Limitations and Discussion}
As a trial to upgrade mesh-based workflow using recent neural techniques, our approach can generate dynamic human assets compatible with existing post-production tools, and enables photo-real human modeling and rendering for streamable immesive experiences. Despite such compelling capabilities, our pipeline still yields to some limitations. Here we provide detailed analysis and discuss potential future extensions.

% 1. rely on background matting
First, our approach relies on image-based greenscreen segmentation to separate the foreground performances, without providing view-consistent matting results. Thus, our approach yields to the segmentation error and still  cannot provide high-quality hair rendering with opacity details. Inspired by recent work~\cite{barron2022mipnerf360,luo2022artemis}, we plan to explore the explicit modeling of background and opacity hairs into our neural framework.
% 2. non-rigid scene with large motions, changing clothes, human-object interactions, when tracing fails 
Moreover, our rendering pipeline relies on the animated meshes and is thus limited by the tracking quality. Even using neural blending from virtual views, our approach still suffers from rendering artifacts when handling extremely fast motions, severe topology changes like removing clothes or complex human-object interactions. It remains an unsolved area of active research towards these scenarios.

% 3. bottleneck for run-time. streaming, not fully end-to-end real-time for live streaming
As shown in the runtime breakdown analysis on Table.~1, our pipeline only supports real-time playback, and still cannot achieve on-the-fly modeling and rendering for real-time live streaming. The bottleneck lies on the deformation and especially the implicit surface learning of Instant-NSR. For acceleration, we point out to adopt finite difference for CUDA-based normal estimation, but such approximated normal will degrade the surface details. A more faithful CUDA implementation requires months of engineering efforts to figure out the backprogation of second-order gradients for both MLP and hash features, which is not fully-supported in existing CUDA libraries or code bases~\cite{mueller2022instant,tiny-cuda-nn}. To stimulate future work towards this direction, we will make our PyTorch implementation publicly available.

% 4. not end-to-end, utilmate solution would be transmit a compact neural representation, but can be super-efficiantly change into various
For the pipeline design, our approach facilitates various neural techniques for reconstruction, tracking and rendering of human performances, respectively, in contrast to recent implicit NeRF-like frameworks~\cite{zhang2021editable,tretschk2020non}. It's promising to further design and transmit a compact neural representation, and then decode it into geometry, motions or appearance to support various downstream applications. We view our approach as a solid step towards this ultimate goal, by intermediately utilizing animated mesh-based workflow and lifting it into the neural era. 
% 5. drivable, relighting
Besides, thanks to the efficiency of our approach to generate dynamic human assets, we plan to scale up our dome-level datasets and explore to generate animatable digital humans~\cite{xiang2021modeling}. It's also interesting to adopt explicit decomposition of light and materials into our approach for relightable human modeling.

	\section{Conclusion}
We have presented a comprehensive neural modeling and rendering approach for high-quality reconstruction, compression, and rendering of human performances from dense multi-view RGB inputs. Our key idea is to lift the traditional animated mesh workflow into the neural era with a new class of highly efficient neural techniques. Our neural surface reconstructor, Instant-NSR efficiently generates high-quality surfaces in minutes, by combing TSDF-based volumetric rendering with hash encoding. Our hybrid tracker compresses the geometry and encodes temporal information by generating animated meshes in a self-supervised manner, supporting various conventional mesh-based rendering using dynamic textures or lumigraph. We further propose a hierarchical neural rendering scheme to strike an intricate balance between quality and bandwidth. We showcase the capabilities of our approach for in a variety of mesh-based applications and photo-real immersive experience with 4D human performance playback in virtual and augmented reality. We believe that our approach serves as a strong transition from well-established mesh-based workflow to neural human modeling and rendering. It makes a solid step forward to faithfully recording the performances of us humans, with numerous potential applications for entertainment, gaming and immersive experience in VR/AR and Metaverse.

	\begin{acks}
		
		% The authors would like to thank Dr. Maura Turolla of Telecom
		% Italia for providing specifications about the application scenario.
		
		% The work is supported by the \grantsponsor{GS501100001809}{National
		%  Natural Science Foundation of
		%  China}{http://dx.doi.org/10.13039/501100001809} under Grant
		% No.:~\grantnum{GS501100001809}{61273304\_a}
		% and~\grantnum[http://www.nnsf.cn/youngscientists]{GS501100001809}{Young
		%  Scientists' Support Program}.
		The authors would like to thank DGene Digital Technology Co., Ltd. for processing the dataset. Besides, we thank Bo Yang from ShanghaiTech University for producing a supplementary video.
		This work was supported by Shanghai YangFan Program (21YF1429500), Shanghai local college capacity building program (22010502800), NSFC programs (61976138, 61977047), the National Key Research and Development Program (2018YFB2100 500), STCSM (2015F0203-000-06) and SOME (2019-01-07-00-01- E00003).

	\end{acks}
	
	\bibliographystyle{ACM-Reference-Format}
	\bibliography{ref}

	\appendix
\section*{APPENDIX}
\setcounter{section}{1}
\subsection{Neural Reconstructor Implementation Details}
\paragraph{Network Architecture.}
In Sec.~3, we introduce a neural surface reconstructor called Instant-NSR, to efficiently achieve high-quality reconstruction results. 
Similar to the network architecture of Instant-NGP, our Instant-NSR consists of two concatenated MLPs: a 2-hidden-layer SDF MLP $m_s$ and a 3-hidden-layer color MLP $ m_c$, both 64 neurons wide. 
In the former SDF MLP, we replace the original ReLU activation with Softplus and set $\beta = 100$ for the activation functions of all the hidden layers.
The SDF MLP maps the 3D position to 32 output values using a hash encoding function [Müller et al. 2022]. Specifically, the input of the SDF MLP is the concatenation of
\begin{itemize} 
	\setlength\itemsep{0em}
	\item the 3 input spatial location values of each 3D sampled point, 
	\item the 32 output values from the hash encoded position.
\end{itemize} 
Then, we apply a truncated function to the output SDF value which maps it to $[-1, 1]$ using the sigmoid activation.

The color MLP also adds view-dependent color variation by spherical harmonics encoding function [Müller et al. 2022]. Its input is the concatenation of
\begin{itemize} 
	\setlength\itemsep{0em}
	\item the 3 input spatial location values of each 3D sampled point,
	
	\item the 3 estimated normal values from the approximated SDF gradient by finite
	difference function,
	
	\item the 16 output values of the SDF MLP, and
	
	\item the view direction decomposed onto the first 16 coefficients of the spherical harmonics basis up to degree 4.
\end{itemize} 
Similar to concurrent work [Mildenhall et al. 2021], we further use a sigmoid activation to map the output RGB color values into the range $[0, 1]$. 

\paragraph{Training details.}
We have demonstrated in the paper that our $\sim$10min training results are comparable with the $\sim$8h optimization results of the original NeuS [Wang et al. 2021a].
In the optimization phase, we assume the region of interest is inside a unit sphere at the very beginning. We also adopt the hierarchical sampling strategy from [Mildenhall et al. 2021] in our PyTorch implementation, where the number of coarse and fine sampling is 64 and 64, respectively. 
We sample $4,096$ rays per batch and train our model for 6k iterations for 12 minutes using a single NVIDIA RTX 3090 GPU.
To approximate the gradient for efficient normal calculation, we adopt the finite difference function $f'(x) = (f(x + \Delta x) - (fx-\Delta x)) / 2\Delta x$ as described in Sec.~3.2. 
In our PyTorch implenmentation, we set approximating step as $\Delta x = 0.005$ and decrease it to $\Delta x = 0.0005$ at the end of training. 
we optimize our model by minimizing both a Huber loss $\mathcal{L} _{color}$ and an Eikonal loss $\mathcal{L} _{eik}$. 
These two losses are balanced with an empirical coefficient $\lambda$, which is set to be $0.1$ in our experiments.
Besides, we choose Adam [Diederik P Kingma et al.2014] optimizer with an initial learning rate of $1e-2$ and decrease it to $1.6e-3$ during training.

\setcounter{section}{1}
\subsection{Neural Tracking Implementation Details}
In Sec.~4, we propose a neural tracking pipeline that marries the traditional non-rigid tracking and neural deformation net in a coarse-to-fine manner. 
We solve the non-rigid tracking via the Gauss-Newton method and introduce details in the following.
\paragraph{Tracking details.}
Before non-rigid tracking, we sampled ED nodes by computing the geodesic distance on the canonical mesh. 
We calculate the average edge length and multiply it by a radius ratio that is used to control the degree of compression to obtain influence radius $r$.
Through all our experiments, we found that simply adjusting to 0.075 also gives good results. 
Given the $r$, we sort the vertices by Y-axis and pick the ED node when it is outside $r$ from the set of existing ED nodes. 
Besides, we can link the ED nodes when they influence the same vertices and then construct the ED graph in advance for subsequent optimization.

\paragraph{Network Architecture.}
The key of our refinement stage includes the canonical radiance field $\phi^o$ and the deformation network $\phi^d$. 
Specifically, $\phi^o$ has the same network architecture as Instant-NGP consisting of the three-dimensional hash encoding and two concatenated MLPs, density and color. 
3-dimensional coordinates are mapped to 64-dimensional features by hash encoding as the input to the density MLP.
Then, the density MLP has a 2-hidden-layer(with 64 hidden dimensions) and outputs a 1-dimension density and 15-dimensions geometry features. 
The geometry features are concatenated with the direction encoding and are fed into the color MLP which has a 3-hidden-layer.
Finally, we can acquire the density value and RGB value per coordinate.
$\phi^d$ includes the four-dimensional hash encoding and a single MLP. Specifically, the four-dimensional hash encoding has 32 hash tables that map the input $(\Vec{p}',t)$ to the 64-dimensional features. Through our 2-hidden-layer deform MLP (with 128 hidden dimensions), we can finally get the $\Delta \Vec{p}'$.

\paragraph{Training details.}
We train the $\phi^o$ and $\phi^d$ separately. 
We first utilize the multi-view images to train the canonical representation $\phi^o$. 
When the PSNR value has stabilized (normally after 100 epochs), we freeze the $\phi^o$ parameters. 
Then, we train the deformation network $\phi^d$ to predict per-frame warping displacement. 
We build a PyTorch CUDA extension library to achieve fast training. We first transform ED nodes from canonical frame to current frame then build a KNN volume. 
Specifically, our volume resolution is $256^3$ to $512^3$ and for each voxel in the KNN volume, we query the 4 to 12 Nearest Neighbor ED nodes via heap. 
Based on the KNN volume, we can quickly query any 3D points in the volume and get the neighbors and the correspondence skinning weights to calculate the coordinates through non-rigid tracking. 

\setcounter{section}{1}
\subsection{Neural Blending Implementation Details}
\paragraph{Network Architecture.}
In Sec.~5, we introduce a neural texture blending method that can efficiently blend fine-grained texture details from adjacent virtual views into the novel view. 
We provide the detailed network architecture specifications of the adopted U-net in Tab.~\ref{blend_net}. 

\paragraph{Training details.}

\begin{table}[h!]
	\centering
	\begin{tabular}{lcccc}
		\toprule
		\textbf{Layer} & \textbf{k} & \textbf{s} & \textbf{chns} & \textbf{Input}\\
		\midrule
		\hline
		Conv1 &  3&  1& 9/32 & RGBD1+RGBD2+Occlusion\\ \hline
		DownSample1 &  3&  2& 32/48& Conv1\\ \hline
		DownSample2 &  3&  2& 48/64& DownSample1\\ \hline
		DownSample3 &  3&  2& 64/96& DownSample2\\ \hline
		DownSample4 &  3&  2& 96/128& DownSample3\\ \hline
		UpSample1 & 3 &  -& 128/96& DownSample4\\ \hline
		UpSample2 & 3 &  -& 192/64& UpSample1+DownSample3\\ \hline
		UpSample3 & 3  &  -& 128/48& UpSample2+DownSample2\\ \hline
		UpSample4 & 3 &  -& 96/32 &  UpSample3+DownSample1\\ \hline
		Weight & 3 & 1 & 32/3 & UpSample4 \\ \hline
		
		\bottomrule
	\end{tabular}
	
	\rule{0pt}{0.01pt}
	\caption{Details about our network architecture of neural blending module. Here, k is the kernel size, s is the stride, d is the kernel dilation, and chns shows the number of input and output channels for each layer, Input means the input of the current layer.}
	\vspace{-0.3cm}
	\label{blend_net}
\end{table}

During the training process, we introduce a new dimension called occlusion map, which is calculated as the difference between two warped depth maps. Then, we utilize the occlusion map (1-dimension) and the two warped RGBD channels (4-dimension) as the network input, which can further help the U-net network to optimize the blending weight.
In the traditional pixel-wise neural texture blending process, the blending result is only generated from the two warped images. However, it will suffer from severe artifacts if the result under the target view is occluded in both adjacent virtual views. 
Thus, we use the texture rendering result as the additional input to recover the missing part due to occlusion. 
For efficiently rendering, we first down-sample the input image to $512\times512$ as the network input, and then up-sample the weight map by bi-linear interpolation to generate the final 2K image.
To avoid the blending network over-fitting to additional texture rendering input, we apply Gaussian blur operation to simulate the low-resolution texture rendering images during the training process. 
This operation helps the network to focus on the details from the selected adjacent views while recovering the missing part from the additional texture rendering input.
Besides, we choose Adam [Diederik P Kingma et al.2014] optimizer with an initial learning rate of $1e-4$ and weight decay rate of $5e-5$. 
We pre-train our neural texture blending model for two days using Twindom [web.twindom.com] dataset on a single NVIDIA RTX 3090 GPU. 
\paragraph{Training datasets.}
To train a more adaptive neural texture blending network, we construct a massive synthetic multi-view dataset.
We utilize the pre-scanned models from Twindom [web.twindom.com]  dataset to generate multi-view images.
Specifically, we re-render the pre-scanned models to generate the six fixed views for blending, and sample 180 virtual target views on a sphere to train the network.
To enhance the generation ability of our network, we enlarge the training dataset by rigging the 3D model to add more challenging poses.

\end{document}